\documentclass[pre,twocolumn,showpacs,floatfix]{revtex4}
\usepackage{graphicx}

\def\be{\begin{equation}}
\def\ee{\end{equation}}

\begin{document}
\title{One-dimensional Langevin models of fluid particle acceleration
in developed turbulence}
 \author{A.K. Aringazin}
 \email{aringazin@mail.kz}
  \altaffiliation[Also at ]
  {Department of Mechanics and Mathematics, Kazakhstan Division, Moscow State
University, Moscow 119899, Russia.}
 \affiliation{Department of Theoretical Physics, Institute for
Basic Research, Eurasian National University, Astana 473021
Kazakhstan}

 \author{M.I. Mazhitov}
 \email{mmi@emu.kz}
 \affiliation{Department of Theoretical Physics, Institute for
Basic Research, Eurasian National University, Astana 473021
Kazakhstan}

\date{18 October 2003}

\begin{abstract}
We make a comparative analysis of some recent one-dimensional
Langevin models of the acceleration of a Lagrangian fluid particle
in developed turbulent flow. The class of models characterized by
random intensities of noises (RIN models) provides a fit to the
recent experimental data on the acceleration statistics. We review
the model by Laval, Dubrulle, and Nazarenko (LDN) formulated in
terms of temporal velocity derivative in the rapid distortion
theory approach, and propose its extension due to the RIN
framework. The fit of the contribution to fourth order moment of
the acceleration is found to be better than in the other
stochastic models.
We study the acceleration probability density function conditional
on velocity fluctuations implied by the RIN approach to LDN type
model. The shapes of the conditional distributions and the
conditional acceleration variance have been found in a good
agreement with the recent experimental data by Mordant, Crawford,
and Bodenschatz (2003).
\end{abstract}


\maketitle

\section{Introduction}\label{Sec:Introduction}

Tsallis statistics \cite{Tsallis} inspired formalism
\cite{Johal,Aringazin,Beck3} was recently used by C.~Beck
\cite{Beck,Beck4} to describe Lagrangian statistical properties of
developed turbulence; see also~\cite{Wilk,Beck2,Reynolds}. In
recent papers~\cite{Aringazin2,Aringazin3,Aringazin4} we have made
some refinements of this approach. The probability density
function of a component of the Lagrangian acceleration of
infinitesimal fluid particle in the developed turbulent flow is
found due to the equation,
\be\label{P}
P(a) = \int_{0}^{\infty}\! d\beta\ P(a|\beta)f(\beta),
\ee
where $P(a|\beta)$ is a conditional probability density function
associated to a surrogate dynamical equation, the one-dimensional
Langevin equation for the acceleration $a$,
\be\label{Langevin}
\partial_t a = \gamma F(a) + \sigma L(t).
\ee
Here, $\partial_t$ denotes time derivative, $F(a)$ is the
deterministic drift force, $\gamma$ is the drift coefficient,
$\sigma^2$ measures intensity of the noise, a strength of the
additive stochastic force, $L(t)$ is Langevin source, a
delta-correlated Gaussian white noise with zero mean,
\be\label{Lnoise}
\langle L(t)\rangle=0, \quad \langle L(t)L(t')\rangle=2\delta(t-t'),
\ee
where the averaging is made over ensemble realizations.

For constant parameters $\gamma$ and $\sigma$, this usual Langevin
model ensures that the stochastic process $a(t)$ defined by
Eq.~(\ref{Langevin}) is Markovian. The probability density
function, $P(a|\beta)$, of the acceleration at fixed $\beta$,
\be\label{beta0}
\beta =\gamma/\sigma^2,
\ee
can be found as a stationary solution of the corresponding
Fokker-Planck equation,
\be\label{FPconstant}
\partial_t P(a,t)
 = \partial_a[-\gamma F(a) + \sigma^2\partial_a]P(a,t),
\ee
where $\partial_a =\partial/\partial a$. This equation can be
derived from the Langevin equation (\ref{Langevin}) using the
noise (\ref{Lnoise}) either in Stratonovich or Ito
interpretations. Particularly, for a linear drift force,
$F(a)=-a$, the stationary probability density function,
$\partial_t P(a,t)=0$, is of a Gaussian form,
\be\label{PGauss}
P(a|\beta)= C(\beta)\exp[-\beta a^2/2],
\ee
where $C(\beta)=\sqrt{\beta/2\pi}$ is a normalization constant,
$a\in [-\infty,\infty]$. The function $f(\beta)$ entering
Eq.~(\ref{P}) is a probability density function arising from the
assumption that $\beta$ is a random parameter with prescribed
external statistics.

While it is evident that the three-dimensional Navier-Stokes
equation with a delta-correlated Gaussian white random forcing
belongs to a class of non-linear stochastic dynamical equations
for the velocity field with which one can associate some
generalized Fokker-Planck equations, it is a theoretical challenge
to make a link between the Navier-Stokes equation and surrogate
one-dimensional Langevin models for acceleration such as
(\ref{Langevin}). This model is, of course, far from being a full
model of the essential Lagrangian dynamics of fluid in the
developed turbulence regime.

Review and critical analysis of the applications of various recent
nonextensive statistics based models to the turbulence have been
made by Gotoh and Kraichnan~\cite{Kraichnan0305040}. An emphasis
was made that some models lack justification of a fit from
turbulence dynamics although being able to reproduce experimental
data to more or less accuracy. A deductive support from the
three-dimensional Navier-Stokes equation was stressed to be
essential for the fitting procedure to be considered meaningful.

In contrast to the fluid particle velocity, the fluid particle
acceleration,
\be\label{Eaccel}
a_i=\frac{dv_i}{dt}\equiv\partial_tv_i+v_k\partial_kv_i,
\ee
which incorporates the Eulerian local acceleration and nonlinear
advection term, can be measured easier by using the Lagrangian
framework while in the Eulerian framework (fixed probe) this
requires measurements of the velocity $v_i$, and temporal and
spatial velocity derivatives, $\partial_tv_i$ and $\partial_kv_i$,
where $\partial_k=\partial/\partial x^k$ denotes spatial
derivative in the Cartesian laboratory frame of reference; $i,k =
1,2,3$. In the Lagrangian framework, the Navier-Stokes equation
can be written as
\be\label{NSaccel}
a_i=-\rho^{-1}\partial_i p + \nu\partial^2_k v_i + f_i,
\ee
where $\rho$ is constant fluid density, $p$ is pressure, $\nu$ is
kinematic viscosity, $v_i=\partial_tx_i$ is velocity, and $f_i$ is
forcing. Here, $x_i=X_i(x_{0k},t)$ is the particle coordinate
viewed as a function of the initial value $x_i(0)=x_{0i}$ and time
$t$ so that the measurement of time series $x_i(t)$ of some
individual particle by using a fine finite-difference scheme
allows one to evaluate its acceleration as a function of time by
using the Lagrangian relation,
\be\label{Laccel}
a_i =\partial^2_tx_i.
\ee
With the initial data points $x_{0i}$ (Lagrangian coordinates)
running over all the fluid particles one gets a Lagrangian
description of the fluid flow. Direct analytical evaluation of the
acceleration from Eq.~(\ref{NSaccel}) is out of reach at present
so that one is led to estimate it in some fashion.

The model (\ref{Langevin}) belongs to a class of stochastic models
of Lagrangian turbulence and deals with an evolution of the
acceleration in time which in accord to the Navier-Stokes equation
is driven by time derivative of the r.h.s. of Eq.~(\ref{NSaccel}).
This type of modelling corresponds to the wellknown universality
(Kolmogorov 1941, Heisenberg 1948, Yaglom 1949) in statistically
homogeneous and isotropic developed turbulence which is expected
to occur in the inertial range only statistically. Accordingly,
the velocity and acceleration become random, and one is interested
in their probability density functions, or multipoint correlation
functions. This is in an agreement with the observed temporally
irregular character of the velocity and acceleration of a tracer
particle in high-Reynolds-number turbulent flows. By the
universality, statistics of the velocity and statistics of the
acceleration do not depend on statistics of the forcing and chosen
initial data. In this paper we are interested in statistics of one
of the acceleration components, $a$, so that we model its
evolution in time.

In a physical context, an essential fluid particle dynamics in the
developed turbulent flow is described here in terms of a
generalized Brownian like motion, a stochastic particle approach,
taking the particle acceleration (\ref{Laccel}) as the dynamical
variable. Such models are generally based upon a hierarchy of
characteristic time scales in the system and naturally employ
one-point statistical description using Langevin type equation (a
stochastic differential equation of first order) for the dynamical
variable, or the associated Fokker-Planck equation (a partial
differential equation) for one-point probability density function.

With the choice of delta-correlated noises such models fall into
the class of Markovian models (no memory effects at small scales)
allowing well established Fokker-Planck approximation. The
consideration of finite-time correlated noises and the associated
memory effects requires a deeper analysis which should be made
separately in each particular case. The evolution equations are
formulated and solved in the Lagrangian framework (the comoving
frame), in a purely temporal treatment, with fluctuations being
treated along the particle trajectory.

Approximation of a short-time correlated noise by the zero-time
correlated one is usually made due to the time scale hierarchy
emerging from the general physical analysis of the system and
experimental data. Under the stationarity condition, a balance
between the energy injected at large scales and the energy
dissipated by viscous processes at small scales, one can try to
solve the Fokker-Planck equation to find stationary probability
density function of the acceleration, $P(a)$. This function as
well as the associated moments can then be compared with the
experimental data on acceleration statistics. The Fokker-Planck
approximation allows one to make a link between the dynamics and
the statistical approach. In the case when stationary probability
distribution can be found exactly one can make a further analysis
without a dynamical reference, yet having a possibility to extract
stationary time correlators.

In contrast to the usual Brownian like motion, the fluid particle
acceleration does not merely follow a random walk with a complete
self-similarity at all scales. It was found to reveal a different,
multiscale self-similarity, which can be seen from wide tails of a
quasi-Gaussian distribution of the experimental probability
distribution $P(a)$. This requires a consideration of some
specific Langevin type equations, which may include nonlinear
terms, {\it e.g.}, to account for turbulent viscosity effect, and
an extension of the usual properties of model forces, additive and
multiplicative noises.

Specifically, the class of models represented by
Eqs.~(\ref{P})-(\ref{PGauss}) is featured by consideration of the
acceleration evolution driven by the "forces" characterized by
fluctuating drift coefficient $\gamma$ (or fluctuating intensity
of multiplicative delta-correlated noise in a more general case)
and/or fluctuating intensity $\sigma^2$ of the additive noise.
This was found to imply stationary distributions of the
acceleration (or velocity increments in time, for finite time
lags) of a quasi-Gaussian form with wide tails which are a
classical signature of the turbulence intermittency, a phenomenon
which developed turbulent flows exhibit at small time scales.
Earlier work on such type of models are due to Castaing, Gagne,
and Hopfinger~\cite{Castaing}, referred to as the Castaing model,
in which a log-normal distribution of fluctuating variance of
intermittent variable was used without reference to a stochastic
dynamical equation.

The difference from the wellknown class of stochastic models with
delta-correlated Gaussian white multiplicative and additive noises
which are also known to imply quasi-Gaussian stationary
distributions with wide tails is that one supposes that {\em
intensities} of the noises are not constant but fluctuate at a
large time scale. We refer to the models with such {Random
Intensities of Noises} as RIN models.

This class of models introduces a two time-scale dynamics, one
associated to a delta-correlation of noises (modelling the
smallest time scale under consideration, usually of the order of
Kolmogorov time) and the other associated to variations of
intensities of the noises, their possible coupling to each other,
and other parameters assumed to occur at large time scales, up to
a few Lagrangian integral times. From a general point of view, one
can assume a hierarchy of a number of characteristic time scales.
However,in the present paper we simplify the consideration in
order to make it more analytically tractable, in accord to the
presence of two characteristic time scales in the Kolmogorov
picture of fully developed turbulence.

In the approximation of two time-scales, one can start with a
Langevin type equation, derive the associated Fokker-Planck
equation in Stratonovich or Ito formulations, and try to find a
stationary solution of the Fokker-Planck equation, in which slowly
fluctuating parameters are taken to be fixed. As the next step,
one evaluates stochastic expectation of the resulting {\em
conditional} probability density function over the parameters with
some distributions assigned to them. By this way one can obtain a
stationary marginal probability density function as the main
prediction of the model.

The dynamical model (\ref{Langevin}) represents a particular
simple one-dimensional RIN model characterized by the presence of
an additive noise (a short time scale) and fluctuating composite
parameter $\beta=\gamma/\sigma^2$ (a long time scale), where
$\gamma$ is simply kinetic coefficient (a multiplicative noise is
not present explicitly) and $\sigma^2$ is the additive noise
intensity.

Two time-scale stochastic dynamics in describing the acceleration
jointly with the velocity and position was used by
Sawford~\cite{Sawford},
\begin{eqnarray}\label{LangevinSawford}
\partial_t a = -(T_L^{-1}+t_\eta^{-1})a + T_L^{-1}t_\eta^{-1}u \\
\nonumber
+ \sqrt{2\sigma_u^2(T_L^{-1}+t_\eta^{-1})
T_L^{-1}t_\eta^{-1}}L(t),
\end{eqnarray}
\be
\partial_t u =a, \quad \partial_t x = u,
\ee
where
\be
T_L= \frac{2\sigma_u^2}{C_0{\bar\epsilon}},
 \quad
t_\eta=\frac{2a_0\nu^{1/2}}{C_0{\bar\epsilon}^{1/2}},
\ee
are two time scales, $T_L\gg t_\eta$, $C_0$, $a_0$ are Lagrangian
structure constants, $\sigma_u^2$ is the variance of the velocity
distribution, and $\bar\epsilon$ is mean energy dissipation rate
per unit mass. This model predicts Gaussian stationary
distributions for the acceleration and velocity reflecting
uncorrelated character of the fluctuations. An obvious extension
of this model is to replace $\bar\epsilon$ by stochastic energy
dissipation rate $\epsilon$, and assume that it is log-normally
distributed in correspondence with the refined Kolmogorov 1962
approach~\cite{K62}.

Recently, an attempt to generalize the Sawford model to the case
of fluctuating parameters has been made by
A.~Reynolds~\cite{Reynolds}, with a very well agreement with the
experimental results being achieved.

The growing interest in studying Langevin type equations to
describe developed turbulence is motivated by the recent high
precision Lagrangian experiments by Porta, Voth, Crawford,
Alexander, and Bodenschatz \cite{Bodenschatz}, the new data by
Crawford, Mordant, Bodenschatz, and Reynolds (the Taylor
microscale Reynolds number is $R_\lambda=690$, the normalized
acceleration range is $[-60,60]\ni a$, the Kolmogorov timescale
$\tau_\eta$ is resolved)~\cite{Bodenschatz2}, Mordant, Delour,
Leveque, Arneodo, and Pinton ($R_\lambda=740$, $a\in [-20,20]$,
$\tau_\eta$ is not resolved) \cite{Mordant}, Mordant, Crawford,
and Bodenschatz \cite{Mordant0303003}, and direct numerical
simulations of the Navier-Stokes equation by Kraichnan and Gotoh
($R_\lambda=380$, $a\in [-150,150]$)~\cite{Gotoh}; the classical
Reynolds number is $\mathrm{Re} = R_\lambda^2/15$. This gives an
important information on the dynamics and new look to the
intermittency in high-Reynolds-number fluid turbulence.

Response characteristics of the polystyrene tracer particle of
about 46 $\mu$m size and the precision in the experiments
\cite{Bodenschatz,Bodenschatz2} allow to resolve about 1/20 of the
Kolmogorov time and 1/20 of the Kolmogorov length in an
$R_\lambda=970$ flow so that the acceleration can be really
resolved, and the particle follows rare violent events within 7\%
of the ideal value of acceleration even at the highest Reynolds
number studied there. For lower Reynolds numbers the resolutions
with respect to Kolmogorov scales are even much higher. The
collected statistics of about 1.7$\times10^8$ data points appeared
to be sufficient to establish finitness of the fourth order moment
of the acceleration, $\langle a^4\rangle$. The acceleration values
are obtained from the measured velocity increments in time by
certain extrapolation to zero time increment, a procedure
requiring handling data points in the dissipative scale
\cite{Bodenschatz,Mordant0303003}.

The stretched exponential fit with three parameters provides a
very well agreement with the experimental data on the transverse
acceleration $a$ of the tracer particle in the $R_\lambda=690$
flow~\cite{Bodenschatz,Mordant0303003},
\be\label{Pexper}
P(a) = C \exp\left[-\frac{a^2}{(1
+\left|{b_1a}/{b_2}\right|^{b_3})b_2^2}\right],
\ee
where $b_1=0.513\pm 0.003$, $b_2= 0.563\pm 0.02$, and $b_3=
1.600\pm 0.003$ are fit parameters, and $C=0.733$ is a
normalization constant. At large acceleration values the tails of
the above $P(a)$ decrease asymptotically as $\exp[-|a|^{0.4}]$,
that implies a convergence of the fourth order moment, $\langle
a^4\rangle =\int_{-\infty}^{\infty} a^4P(a)da$. The flatness
factor of the distribution (\ref{Pexper}) which characterizes the
widening of its tails (when compared with a Gaussian) is
\be\label{flatness}
F\equiv \frac{\langle a^4\rangle}{\langle a^2\rangle^2} \simeq
55.1,
\ee
which should be compared with the flatness of the experimental
curve, $F= 55 \pm 8$~~\cite{Mordant0303003}. We remind that for a
Gaussian distribution $F=3$.

With constant $\beta$, the Gaussian probability density function
(\ref{PGauss}) corresponds to the non-intermittent Kolmogorov 1941
picture of fully developed turbulence, and agrees with the
experimental statistics of components of velocity increments in
time for large time scales, up to the integral time scale.
However, it fails to describe observed Reynolds number dependent
stretched exponential tails of the experimental acceleration
probability density function (\ref{Pexper}) that correspond to
anomalously high probabilities for the tracer particle to have
extremely high accelerations, bursts with dozens of
root-mean-square ({\em rms}) acceleration, in the developed
turbulent flow. Such a high probability of the extreme
acceleration magnitudes is associated to the Lagrangian turbulence
intermittency, which was found to be considerably stronger than
the Eulerian one. Equivalently, one can say that it is related to
an increase of the probability to have large velocity increments
in time with a decrease of the time scale, down to the Kolmogorov
time scale (a statistical viewpoint).

In the Eulerian framework, the turbulence intermittency is usually
understood differently, as an increase of the probability to have
large longitudinal velocity differences at short spatial scales,
and studied through nonlinearity in scaling exponents of velocity
structure functions (a structural viewpoint).

Intermittency of the stochastic energy dissipation rate is related
to the dynamical intermittency of chaoticity in the system that
makes a link between the Eulerian and Lagrangian intermittency to
which we refer below.

The averaging (\ref{P}) of the Gaussian distribution
(\ref{PGauss}) over randomly distributed $\beta$, an evaluation of
the stochastic expectation, was found to be a simple ad hoc
procedure to obtain observable predictions, with one free
parameter, that meet experimental statistical data on the
acceleration of the tracer particle. One can think of this as the
averaging over a large time span for one tracer particle, or as
the averaging over an ensemble of tracer particles, moving in the
three-dimensional flow characterized by random spatially
distributed domains with different values of $\beta$.

Physically one would like to know the processes underlying the
random character of the model parameter $\beta$. Due to the
definition (\ref{beta0}) the random character of $\beta$ is
attributed to a random character of the drift coefficient $\gamma$
and/or the additive noise intensity $\sigma^2$.

The distribution of $\beta$ is not fixed uniquely by the theory so
that a judicious choice of $f(\beta)$ makes a problem in the RIN
model (\ref{P})-(\ref{PGauss}).

The primary aim of the present work is to provide a critical
evaluation of the Langevin modeling approach by making a
comparative analysis of different types of models on the basis of
recent Lagrangian experimental data.

The crucial point is to make a link between the Langevin type
equation and the Navier-Stokes equation. This includes
determination of statistical properties of stochastic terms and
the functional form of deterministic terms, as well as their
dependence on the parameters entering the Navier-Stokes equation
justified for the inertial range of fully developed turbulence.
Also, some extension of the stochastic equation may be required to
account for dependence of the parameters on Lagrangian velocity
fluctuations, in the spirit of second-order stochastic
models~\cite{Sawford} and in correspondence to the Navier-Stokes
equation as the pressure gradient term in the Eulerian framework
can be expressed in terms of the velocity owing to the
incompessibility condition. Strong and nonlocal character of
Lagrangian particle coupling due to pressure effects makes the
main obstacle to derive turbulence statistics from the
Navier-Stokes equation.

The layout of the paper is as follows.

In Sec.~\ref{Sec:Langevinmodels} we review some recent
one-dimensional Langevin models of the developed turbulence.

In Sec.~\ref{Sec:RINmodels}, we outline implications of the Random
Intensity of Noises (RIN) models with the underlying chi-square
(Sec.~\ref{Sec:ChisquareDistribution}) and log-normal
(Sec.~\ref{Sec:LognormalDistribution}) distributions of $\beta$
\cite{Beck,Beck4,Aringazin3}. We review results of a recent
approach \cite{Aringazin4} to specify $f(\beta)$ which is based
upon relating $\beta$ to velocity fluctuations $u$ and using
normal distribution of velocity fluctuations with zero mean
(Sec.~\ref{Sec:GaussianDistribution}). This enables to reproduce
chi-square and log-normal distributions of $\beta$ as particular
cases. In general, this approach assumes that parameters of the
model, such as the intensity of additive noise, depend on velocity
fluctuations, in an agreement with the Heisenberg-Yaglom picture
of developed turbulence.

A nonlinear Langevin and the associated Fokker-Planck equations
obtained by a direct requirement that the probability distribution
satisfies some model-independent scaling relation have been
recently proposed by Hnat, Chapman, and Rowlands~\cite{Hnat} to
describe the measured time series of the solar wind bulk plasma
parameters. We find this result relevant to fluid turbulence since
it is based on a stochastic dynamical framework and leads to the
stationary distribution with exponentially truncated power law
tails, similar to that obtained in the above mentioned RIN models
(Sec.~\ref{Sec:HCRmodel}).

The above one-dimensional Langevin toy models of Lagrangian
turbulence all suffer from the lack of physical interpretation,
{\it e.g.,} of short term dynamics, or small-scale and large-scale
contributions, in the context of three-dimensional Navier-Stokes
equation.

The Navier-Stokes equation based approach to describe statistical
properties of small scale velocity increments, both in the
Eulerian and Lagrangian frames, was developed in much detail by
Laval, Dubrulle, and Nazarenko~\cite{Laval}; see also recent
work~\cite{Laval2}. This approach introduces nonlocal interactions
between well separated large and small scales, elongated triads,
and is referred to as the Rapid Distortion Theory (RDT) approach.
This approach is contrasted with the Gledzer-Ohkitani-Yamada shell
model,
in which interactions of a shell of wave numbers with only its
nearest and next-nearest shells are taken into account. We outline
results of this approach and focus on the proposed
{one-dimensional} Langevin model of Lagrangian turbulence to which
we refer as Laval-Dubrulle-Nazarenko (LDN) model
(Sec.~\ref{Sec:LDNmodel}). Particularly, we calculate exactly the
probability density function of acceleration stemming from this
model.

In Sec.~\ref{Sec:Comparison} we make qualitative and quantitative
comparative analysis of the one-dimensional LDN and simple RIN
models.


In Sec.~\ref{Sec:ConditionalProbability} we study conditional
probability density function $P(a|u)$ taking the LDN model with
delta-correlated noises and assuming that the additive noise
intensity parameter $\alpha$ depends on the amplitude of velocity
fluctuations $u$.

\section{One-dimensional Langevin models of the Lagrangian turbulence}
\label{Sec:Langevinmodels}

In this Section, we outline results of some recent one-dimensional
Langevin models of the Lagrangian fluid particle acceleration in
the developed turbulent flow.

\subsection{Simple RIN models}
\label{Sec:RINmodels}

\subsubsection{The underlying chi-square distribution}
\label{Sec:ChisquareDistribution}

With the underlying gamma (chi-square) distribution of $\beta$ of
order $n$ ($n=1,2,3,\dots$),
\be\label{chi2}
f(\beta) =\frac{1}{\Gamma(\frac{n}{2})}
 \left(\frac{n}{2\beta_0}\right)^{\frac{n}{2}}
 \beta^{\frac{n}{2}-1}\exp\left[-\frac{n\beta}{2\beta_0}\right],
\ee
the resulting marginal probability density function (\ref{P}) with
$P(a|\beta)$ given by the Gaussian (\ref{PGauss}) is found in the
form~(cf.~\cite{Beck})
\be\label{Pgamma}
P(a)=\frac{C}{(a^2+{n}/{\beta_0})^{(n+1)/2}},
\ee
where
\be
C= \frac{(n/\beta_0)^{n/2}\Gamma(\frac{n+1}{2})}
{\sqrt{\pi}\Gamma(\frac{n}{2})}
\ee
is normalization constant. With $n=3$  ($\beta_0=3$ for a unit
variance) one obtains the normalized marginal distribution in the
following simple form,
\be\label{Pchi2}
P(a)=\frac{2}{\pi(a^2+1)^2},
\ee
a prediction of the chi-square model, with the Tsallis entropic
index  taken to be $q=(n+3)/(n+1)=3/2$ due to the theoretical
argument that the number of independent random variables at
Kolmogorov scale is $n=3$ for the three-dimensional
flow~\cite{Beck}. One can see that the resulting marginal
distribution is characterized by power law tails that {\it a
priori} lead to divergent higher moments.

A Gaussian truncation of the power law tails naturally arises
under the assumption that the parameter $\beta$ contains a
non-fluctuating part, which can be separated out as follows:
$\beta/2 \to a_c^{-2} +\beta/2$~\cite{Aringazin3}. This leads to
the modified marginal distribution,
\be\label{PAringazin}
P(a) = \frac{C\exp[-a^2/a_c^2]}{(a^2+{n}/{\beta_0})^{(n+1)/2}},
\ee
where $C$ is normalization constant and $a_c>0$ is a free
parameter which can be used for a fitting. Taking the theoretical
value $n=3$ and $\beta_0=3$ as in the above case, one obtains that
\be
C= \frac{2a_c^2}{\pi
(a_c^2-2)\exp[a_c^{-2}](1-\mathrm{erf}(a_c^{-1}))+2\sqrt{\pi}a_c},
\ee
where $\mathrm{erf}(x)$ denotes the error function. The
distribution (\ref{PAringazin}) at the fitted value $a_c=39.0$ is
in a good agreement with the experimental probability density
function $P(a)$~\cite{Bodenschatz,Bodenschatz2}.

Note that at $a_c \to \infty$ (no constant part) the model
(\ref{PAringazin}) covers the model (\ref{Pgamma}). Within the
framework of Tsallis nonextensive statistics, the parameter $q-1$
measures a variance of fluctuations. For $q\to 1$ (no
fluctuations), Eq.~(\ref{PAringazin}) reduces to a Gaussian
distribution, which meets the experimental data for temporal
velocity increments at the integral time scale.


\begin{figure}[tbp!]
\begin{center}
\includegraphics[width=0.45\textwidth]{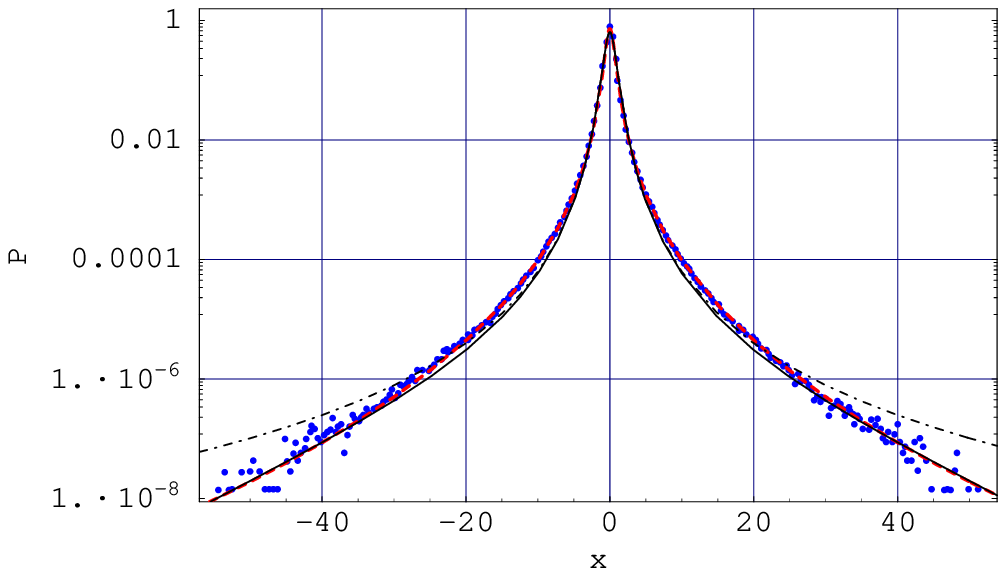}
\includegraphics[width=0.45\textwidth]{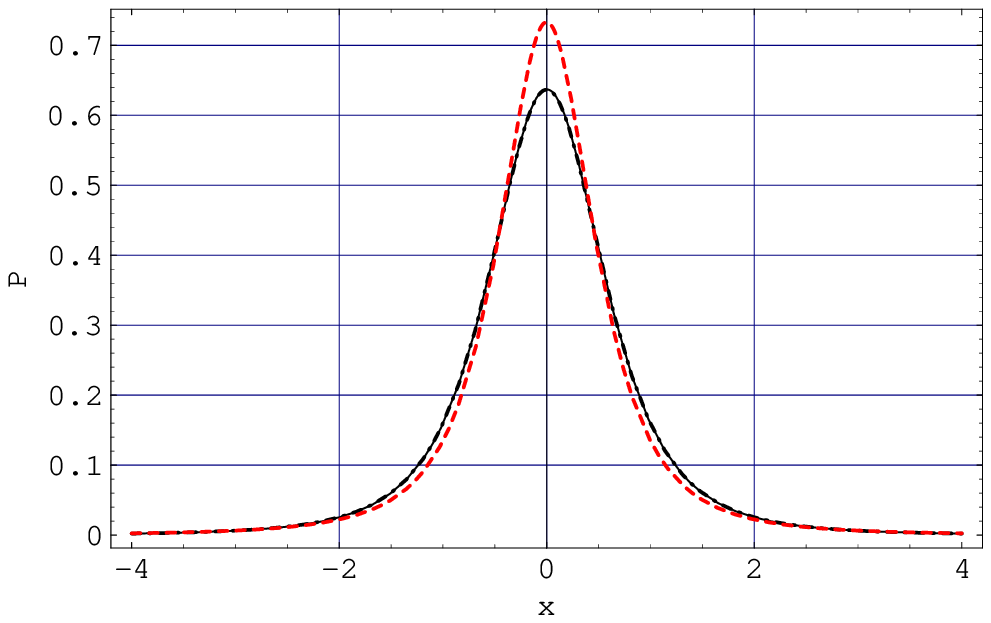}
\end{center}
\caption{ \label{Fig1} Acceleration probability density function
$P(a)$. Dots: experimental data at $R_\lambda=690$ by Crawford,
Mordant, and Bodenschatz~\cite{Bodenschatz2}.  Dashed line:
stretched exponential fit (\ref{Pexper}), $b_1=0.513$, $b_2=
0.563$, $b_3= 1.600$, $C=0.733$. Dot-dashed line: Beck chi-square
model (\ref{Pchi2}), $q=3/2$. Solid line: Chi-square Gaussian
model (\ref{PAringazin}), $a_c=39.0$, $C=0.637$. $x=a/\langle a^2
\rangle^{1/2}$ denotes normalized acceleration.}
\end{figure}

\begin{figure}[tbp!]
\begin{center}
\includegraphics[width=0.45\textwidth]{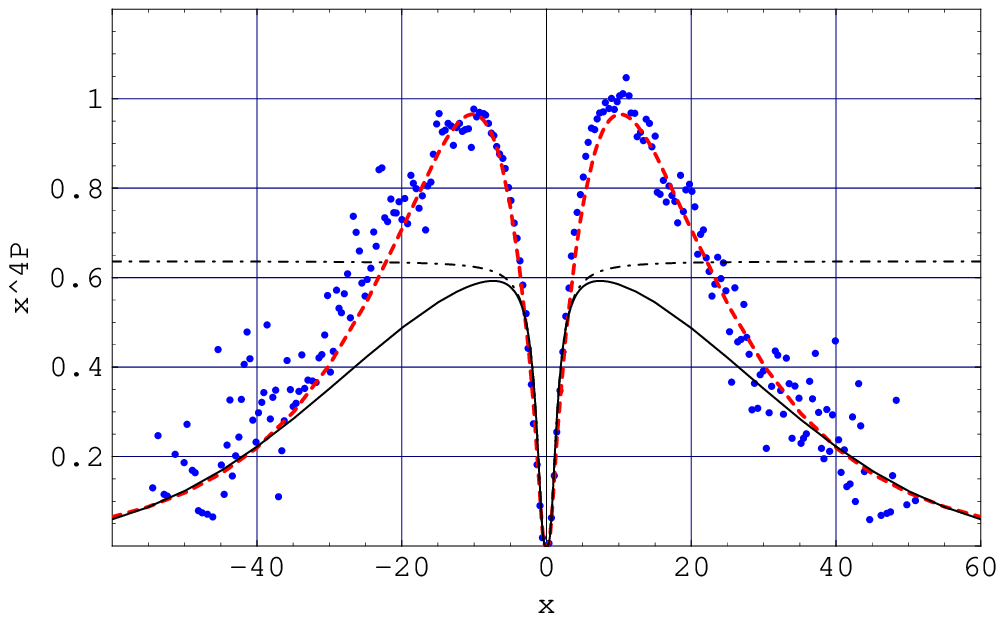}
\includegraphics[width=0.45\textwidth]{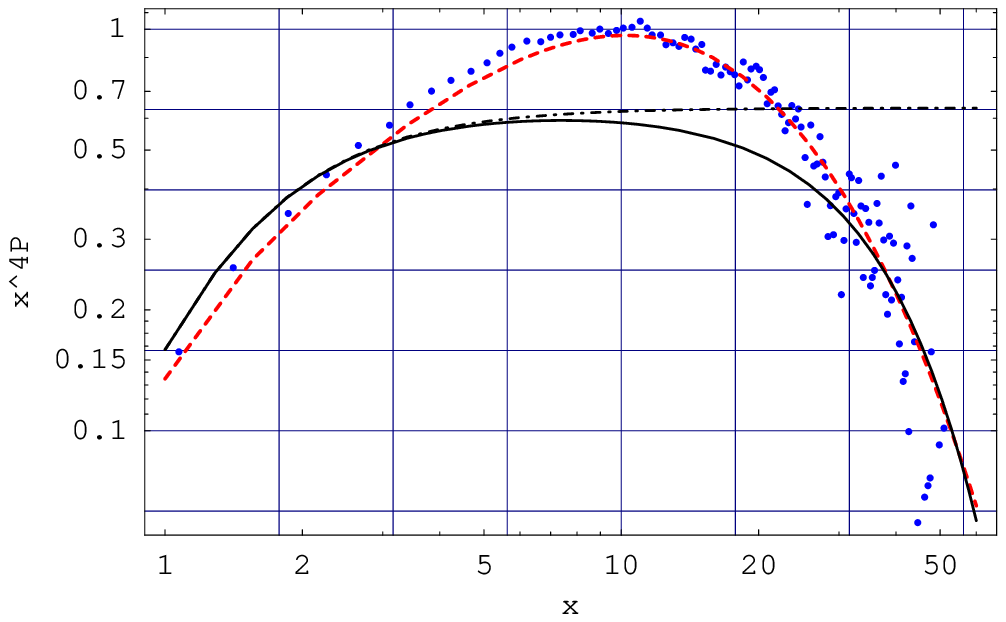}
\end{center}
\caption{ \label{Fig2} Contribution to the fourth order moment
$a^4P(a)$. Top panel: a linear plot, bottom panel: a log-log plot.
Notation is the same as in Fig.~\ref{Fig1}.}
\end{figure}

A comparison of the chi-square model (\ref{Pchi2}) and chi-square
Gaussian model (\ref{PAringazin}) with the experimental data is
shown in Figs.~\ref{Fig1} and \ref{Fig2}. One can see that both
the distributions follow the experimental $P(a)$ to a good
accuracy (at least up to 3 standard deviations), although the
tails of the chi-square model distribution departure to the
experimental curve at big $|a|$. A major difference is seen from
the contribution to the fourth order moment $a^4P(a)$ shown in
Fig.~\ref{Fig2}. The chi-square model yields a qualitatively
unsatisfactory behavior indicating a divergency of the predicted
fourth order moment. In contrast, the chi-square Gaussian model is
in a good qualitative agreement with the data, reproducing them
well at small and large acceleration values although
quantitatively it deviates at intermediate acceleration values and
gives the flatness value $F \simeq 46.1$ for $a_c=39.0$, as
compared to the flatness value $(\ref{flatness})$.

\subsubsection{The underlying log-normal distribution}
\label{Sec:LognormalDistribution}

\begin{figure}[tbp!]
\begin{center}
\includegraphics[width=0.45\textwidth]{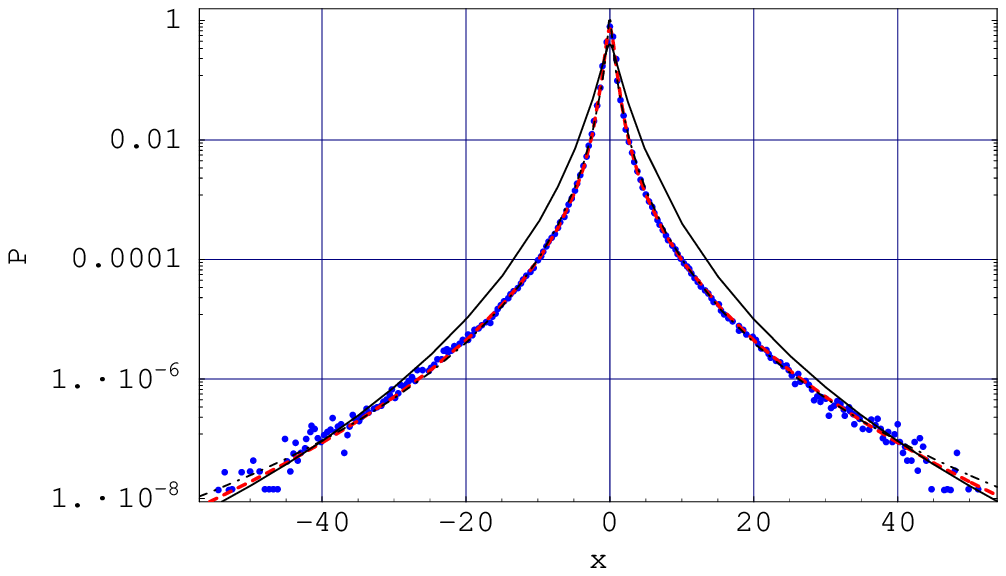}
\includegraphics[width=0.45\textwidth]{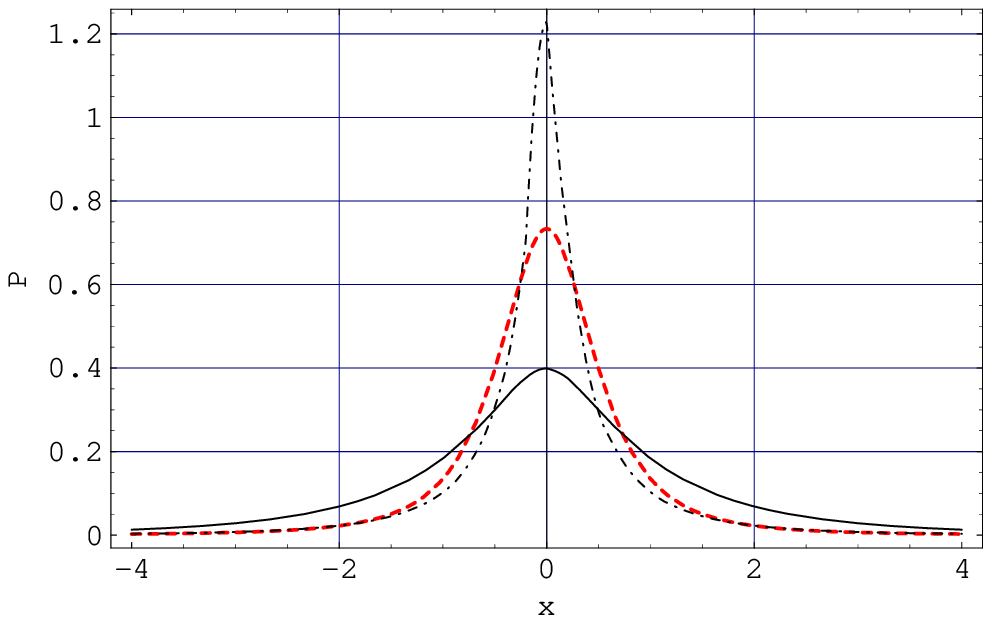}
\end{center}
\caption{ \label{Fig3} Acceleration probability density function
$P(a)$. Dots: experimental data at $R_\lambda=690$ by Crawford,
Mordant, and Bodenschatz~\cite{Bodenschatz2}.  Dashed line:
stretched exponential fit (\ref{Pexper}), $b_1=0.513$, $b_2=
0.563$, $b_3= 1.600$, $C=0.733$. Dot-dashed line: Beck log-normal
model (\ref{PBeck}), $s=3.0$. Solid line: Castaing log-normal
model (\ref{PCastaing}), $s_0=0.625$. $x=a/\langle a^2
\rangle^{1/2}$ denotes normalized acceleration.}
\end{figure}

\begin{figure}[tbp!]
\begin{center}
\includegraphics[width=0.45\textwidth]{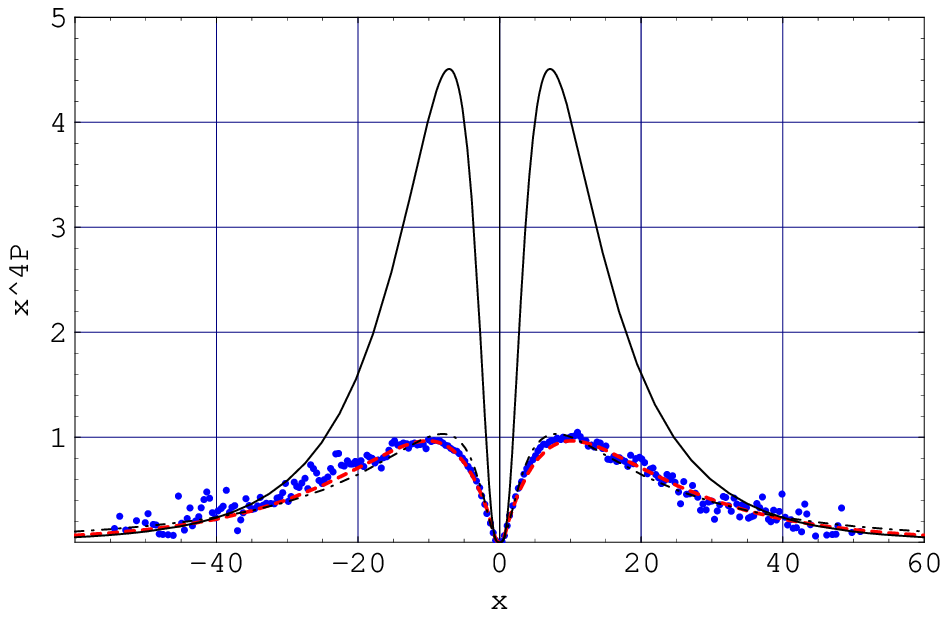}
\includegraphics[width=0.45\textwidth]{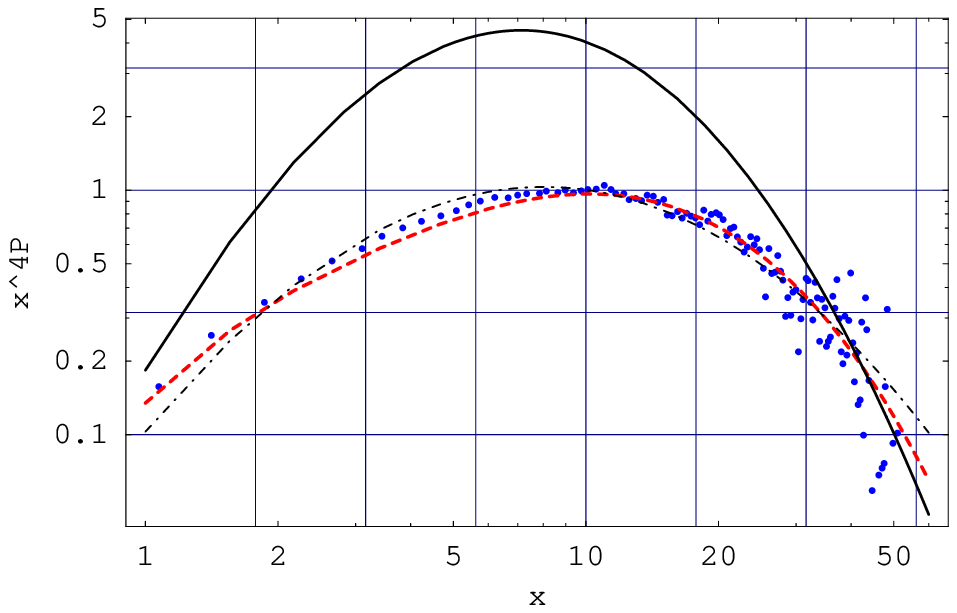}
\end{center}
\caption{ \label{Fig4} Contribution to the fourth order moment
$a^4P(a)$. Top panel: a linear plot, bottom panel: a log-log plot.
Notation is the same as in Fig.~\ref{Fig3}.}
\end{figure}

With the underlying log-normal distribution of $\beta$,
\be\label{lognormal}
f(\beta) =
\frac{1}{\sqrt{2\pi}s\beta}
\exp\left[-\frac{(\ln\frac{\beta}{m})^2}{2s^2}\right],
\ee
the resulting marginal probability density function (\ref{P}) with
$P(a|\beta)$ given by the Gaussian (\ref{PGauss}) was recently
proposed to be \cite{Beck4}
\be\label{PBeck}
P(a) = \frac{1}{2\pi s}\int_{0}^{\infty}\!\!\! d\beta\
\beta^{-1/2}
\exp\left[-\frac{(\ln\frac{\beta}{m})^2}{2s^2}\right]
e^{-\frac{1}{2}\beta a^2},
\ee
where the only free parameter, $s$, can be used for a fitting, or
derived from theoretical arguments, $s^2=3$ ($m=\exp[s^2/2]$ for a
unit variance). This distribution is shown in Fig.~\ref{Fig3} and
was found to be in a good agreement with the Lagrangian
experimental data by Porta {\it et al.}~\cite{Bodenschatz}, the
new data by Crawford {\it et al.}~\cite{Bodenschatz2}, Mordant
{\it et al.}~\cite{Mordant}, and direct numerical simulations of
the Navier-Stokes equation (DNS) by Kraichnan and
Gotoh~\cite{Gotoh}.

However, the central part of the distribution shown in the bottom
panel of Fig.~\ref{Fig3} reveals greater inaccuracy of the
log-normal model ($P(0)\simeq 1.23$) as compared with that of both
the chi-square and chi-square Gaussian models ($P(0)\simeq 0.65$)
which are almost not distinguishable in the region $|a|/\langle
a^2\rangle^{1/2}\leq 4$ (the bottom panel of Fig.~\ref{Fig1}); see
also recent work by Gotoh and Kraichnan~\cite{Kraichnan0305040}.
This is the main failing of the log-normal model (\ref{PBeck}) for
$s^2=3.0$ although the predicted distribution follows the measured
low probability tails, which are related to the Lagrangian
intermittency, to a good accuracy. The central region of the
experimental curve (\ref{Pexper}) ($P(0)\simeq 0.73$) contains
most weight of the experimental distribution and is the most
accurate part of it, with the relative uncertainty of about 3\%
for $|a|/\langle a^2\rangle^{1/2}<10$ and more than 40\% for
$|a|/\langle a^2\rangle^{1/2}>40$~\cite{Mordant0303003}.

The distribution (\ref{PBeck}) is characterized by a bit bigger
flatness value, $F=3\exp[s^2]\simeq 60.3$ for $s^2=3$, as compared
to the flatness value (\ref{flatness}) which is nevertheless
acceptable from the experimental point of view. The peaks of the
contribution to fourth order moment shown in Fig.~\ref{Fig4} do
not match that of the experimental curve for the $R_\lambda=690$
flow. We note that the best fit is achieved for $s^2$ close to the
theoretical value 3 but this does not significantly improve
overlapping of the peaks with the data points.

One naturally expects that a better correspondence to the
experiment may be achieved by an accounting for small scale
interactions via turbulent viscosity (certain nonlinearity in the
first term of the r.h.s. of Eq.~(\ref{Langevin})) as it implies a
damping of the large events, {\it i.e.,} less pronounced
enhancement of the tails of $P(a)$.

It should be noted that the idea to describe turbulence
intermittency via averaging of the Gaussian distribution over
log-normally distributed variance of some intermittent variable
was proposed long time ago by Castaing, Gagne, and Hopfinger
\cite{Castaing},
\be\label{PCastaing}
P(x) =\frac{1}{2\pi s_0} \int_0^\infty\!\!\!
d\theta\,\theta^{-2}\exp
\left[-\frac{(\ln\frac{\theta}{m_0})^2}{2s_0^2}\right]
e^{-x^2/(2\theta^2)},
\ee
where $x$ is a variable under study. Below, we apply this model to
the Lagrangian acceleration, $x=a$.

In technical terms, the difference from the Castaing log-normal
model is that in Eq.~(\ref{PBeck}) the {\em inverse} square of the
variance, $\beta=\theta^{-2}$, is taken to be log-normally
distributed. In essence, the models (\ref{PBeck}) and
(\ref{PCastaing}) are of the same type, with different parameters
assumed to be fluctuating at a large time scale and hence
different resulting marginal distributions.

One can check that the change of variable, $\theta=\beta^{-1/2}$,
in Eq.~(\ref{PCastaing}) leads to the density function different
from that given by Eq.~(\ref{PBeck}),
\be\label{PCastaing2}
P(x) \simeq \int_0^\infty\!\!\! d\beta\,\beta^{-3/2}\exp
\left[-\frac{(\ln\frac{\beta}{m_1})^2}{2s_1^2}\right]
e^{-\frac{1}{2}\beta x^2},
\ee
where we have denoted $m_1=m_0^{-2}$ and $s_1=2s_0$. Therefore,
the distributions (\ref{PBeck}) and (\ref{PCastaing}) are indeed
not equivalent to each other, both being of a stretched
exponential form.

As to a comparison of the fits, we found that the fit of the
Castaing log-normal model (\ref{PCastaing}) for the acceleration,
with the fitted value $s_0=0.625$ ($m_0=\exp[s_0^2/2]$ for unit
variance), is of a considerably less quality as one can see from
Fig.~\ref{Fig3} and, more clearly, from Fig.~\ref{Fig4}. Positions
of the peaks of $a^4P(a)$ are approximately the same for both the
models, namely, $|a|/\langle a^2\rangle^{1/2}\simeq 8$ as compared
to $|a|/\langle a^2\rangle^{1/2} \simeq 10.2$ for the experimental
curve.

We conclude this Section with the following remark. The Langevin
model of the type (\ref{Langevin}), Fokker-Planck approximation of
the type (\ref{FPconstant}), and the underlying lognormal
distribution (\ref{lognormal}) within the Castaing approach were
recently used by Hnat, Chapman, and Rowlands \cite{Hnat} to
describe intermittency and scaling of the solar wind bulk plasma
parameters.

\subsubsection{The underlying Gaussian distribution of velocity
fluctuations}\label{Sec:GaussianDistribution}

The problem of selecting appropriate distribution of the parameter
$\beta$ among possible ones was recently addressed in
Ref.~\cite{Aringazin4}. A specific model based on the assumption
that the velocity fluctuation $u$ follows normal distribution with
zero mean and variance $s$ was developed. The result is that a
class of underlying distributions of $\beta$ can be encoded in the
function $\beta=\beta(u)$, and the marginal distribution is found
to be
\be\label{Pnormal}
P(a) \!=\! C(s)\!\!\int_{0}^{\infty}\!\!\!\!\!\! d\beta\,
P(a|\beta) \exp\!\left[{-\frac{[u(\beta)]^2}{2s^2}}\right]\!
\left|\frac{du}{d\beta}\right|,
\ee
where $u(\beta)$ is the inverse function. Note that only an {\em
absolute} value of $u$ contributes this probability distribution.
Particularly, the exponential dependence,
\be
\beta(u) = \exp[\pm u],
\ee
features the log-normal distribution of $\beta$ so that
Eq.~(\ref{Pnormal}) leads to Eq.~(\ref{PBeck}) used in
Ref.~\cite{Beck4}, while the $\chi^2$ distribution of order one is
recovered with
\be
\beta(u) = u^2.
\ee

In general, this model is relevant when $\beta(u)$ is a monotonic
Borel function of the stochastic variable $u$ mapping $
[-\infty,\infty] \ni u$ to $[0, \infty] \ni\beta$, and allows one
to rule out some {\it ad hoc} distributions of $\beta$ as well as
to make appropriate generalizations of both $\chi^2$ and
log-normal distributions of the parameter $\beta$.

Below we develop a possible dynamical foundation of the above
model.

The stationary distribution (\ref{Pnormal}) with
$\beta(u)=\exp[\pm u]$ can be associated to the Langevin equation
of the form~\cite{Aringazin4},
\be\label{LangevinExp}
\partial_t a = \gamma F(a) + e^{\omega}L(t),
\ee
where we denote $\omega=\mp u/2$, $u$ follows Gaussian
distribution with zero mean, and we take $\gamma$ = const to
simplify the consideration.

In general, we adopt a viewpoint that statistical properties of
the acceleration $a$ are associated to velocity fluctuations
statistics due to the wellknown Heisenberg-Yaglom theory. This
theory predicts the following scaling of a component of the
acceleration variance,
\be\label{HYscaling}
\langle a^2\rangle
= a_0{\bar u}^{9/2}\nu^{-1/2}L^{-3/2},
\ee
where $\bar u$ is the {\em rms} velocity, $L$ is the integral
scale length, and $a_0$ is the Kolmogorov constant. This
longstanding universal ${\bar u}^{9/2}$ scaling was confirmed by
the recent Lagrangian experiments~\cite{Bodenschatz} to a very
high accuracy, for about seven orders of magnitude in the
acceleration variance, or two orders of the {\em rms} velocity, at
$R_\lambda>500$. At lower Reynolds numbers, $R_\lambda<500$, it
appeared that the Heisenberg-Yaglom scaling significantly deviates
from the experimental data due to the emerging dependence of $a_0$
on $R_\lambda$ (the acceleration is increasingly coupled to large
scales of the flow at low Reynolds numbers).

We note also that in the Sawford model the Langevin equation
(\ref{LangevinSawford}) for $a$ includes velocity fluctuations $u$
and the variance of the velocity distribution.

Below we outline a relationship of the model (\ref{LangevinExp})
to some recent approaches in studying the intermittency.

(i) The form of the last term in Eq.~(\ref{LangevinExp}), in which
$\omega$ can be viewed as a Gaussian process, $\omega=\omega(t)$,
independent on the white noise $L(t)$, strikingly resembles that
involved in the recently developed log-infinitely divisible
multifractal random walks (MRW) model by Muzy and Bacry
\cite{Muzy}, a continuous extension of discrete cascades.

(ii) The driving force amplitude of the form $e^{\omega(t)}$, with
an ultraslow decaying correlation function,
$\langle\omega(t)\omega(t+\tau)\rangle_t =
-\lambda_0^2\ln[\tau/T_L]$, $\tau < T_L$, in Langevin type
equation has been recently considered by Mordant {\it et al.}
\cite{Mordant}; $T_L$ stands for the Lagrangian integral time. The
results of this model have been found in a very good agreement
with the experimentally observed very slow decay of the
equal-position time autocorrelation of the fluid particle velocity
increment magnitude in time, $|\Delta_\tau u_i|$, for each
component (very much like to MRW model) attributed to the
intermittency of Lagrangian trajectories in the developed
turbulent flow. Also, very slow decay was observed for the cross
correlation of the magnitudes of the acceleration components. Both
the dynamical correlations were found to vanish only for
$\tau>3T_L$, while the dynamical correlations of the full signed
entities, $\Delta_\tau u_i$, decay rapidly, the autocorrelation
functions cross zero at about $\tau \simeq 0.06T_L$ and the cross
correlation functions are approximately zero.

We note that the fitted value of the above intermittency parameter
$\lambda_0$ ($\lambda_0^2=0.115\pm 0.01$) is very close to
$1/s^2=1/3$, with the value $s^2=3$ in Eq.~(\ref{PBeck})
interpreted as the number of independent random variables in
three-dimensional space at the Kolmogorov scale. If this is not
due to a coincidence, the intermittency parameter $\lambda_0$
approaches simply the inverse of the effective space dimension
number $d$,
\be
\lambda_0 = 1/d,
\ee
$d=3$, for high-Reynolds-number turbulent flows.

The above connection to the MRW model and very slow decay of the
correlations of the absolute values of acceleration components
indicate relevance of the specific representation
(\ref{LangevinExp}), with very slow varying $|u|$, in the
description of intermittency. In fact, due to the experiments
\cite{Pinton} the Lagrangian velocity autocorrelatation function
$\langle u(t)u(t+\tau)\rangle_t/\langle u^2\rangle$ decays almost
exponentially but very slowly, to vanish only for $\tau > 3T_L$,
where the integral time scale $T_L=2.2\times 10^{-2}$~s, which is
two orders of magnitude bigger than the Kolmogorov time scale,
$\tau_\eta = 2.0\times10^{-4}$~s; $R_\lambda = 740$ and the mean
velocity is about 10\% of the {\it rms} velocity.

(iii) Due to the wellknown Kolmogorov power law scaling
relationship between $\bar\epsilon$ and $\bar u$, the
representation (\ref{LangevinExp}) can be thought of as the result
of use of the relation $\ln\beta \simeq\ln\epsilon$, with
$\ln\epsilon$ being normally distributed due to the refined
Kolmogorov 1962 theory. Here, $\epsilon$ denotes the stochastic
energy dissipation rate per unit mass treated in the Lagrangian
framework. From this point of view, one can identify
\be\label{omega}
\omega = g\ln\epsilon,
\ee
where $g$ is a constant. This means that the stochastic dynamics
of the logarithm of the energy dissipation is independent, and it
influences the acceleration dynamics specifically through the
intensity of driving stochastic force in Eq.~(\ref{LangevinExp}).
Stationary normal distribution of $\omega$ can be in turn derived
from the Fokker-Planck equation associated to the Langevin
equation of a linear form,
\be\label{Langevinomega}
\partial_t \omega = g_0 + g_1 \omega + g_2 L(t),
\ee
where $g_i$ are constants. This equation is in an agreement with
the recent results of Eulerian (hotwire anemometer) study of the
interaction between velocity increments and normalized energy
dissipation rate by Renner, Peinke, and
Friedrich~\cite{Friedrich}. Particularly, they found that an
exponential dependence of the diffusion coefficient on the
logarithmic energy dissipation in the Fokker-Planck equation for
the velocity increments in space is in a very good agreement with
the experimental data. We note that Eq.~(\ref{Langevinomega}) does
not imply a logarithmic decay of the Lagrangian correlation
function $\langle\omega(t)\omega(t+\tau)\rangle$ proposed in
Ref.~\cite{Mordant}. This may be attributed to the wellknown
difference between the Eulerian (fixed probe) and Lagrangian
(trajectory) frameworks.

(iv) For the choice $\beta(u)=\exp[u]$ corresponding to the
log-normal distribution of $\beta$, using Eq.~(\ref{LangevinExp})
one can derive the stationary probability density function of the
form~\cite{Aringazin4}
\be\label{gu}
P(a)= \int_{-\infty}^{\infty}du\ C(u)\exp[\ln[g(u)]-e^{u} a^2/2],
\ee
where $g(u)$ is a probability density function of $u$. Hence the
joint probability density function can be written
\be\label{PDFfluct}
P(a,u)= C(u)\exp[\ln[g(u)]-e^{u}a^2/2].
\ee
Such a form of the distribution, containing specifically the
double exponent, resembles the "universal" distribution of
fluctuations (Gumbel function),
\be\label{PChapman}
P(x)= c_0\exp[c_1(y-e^{y})], \quad y\equiv c_2(x-c_3),
\ee
where $c_i$ are constant, recently considered by Chapman,
Rowlands, and Watkins~\cite{Chapman} (see also references therein)
following the work by Portelli, Holdsworth, and
Pinton~\cite{Portelli}. They used an apparently different approach
(not related to a Langevin type equation for the acceleration
studied in our paper) based on the multifractal type energy
cascade and $\chi^2$ or log-normal (K62 theory) underlying
distribution for the energy dissipation rate at fixed level. They
pointed out a good agreement of such $P(x)$ with experimental
data, where $x$ denotes a fluctuating entity observed in a variety
of model correlated systems, such as turbulence, forest fires, and
sandpiles. The result of this approach meets ours and we consider
it as an alternative way to derive the characteristic probability
measure of fluctuations; with $g(u)$ taken to be a $\chi^2$
(respectively, Gaussian) density function, one obtains, up to a
pre-exponential factor and constants, $P(u)\sim \exp[-u-\exp[u]]$
(respectively, $P(u)\sim \exp[-u^2-\exp[u]]$). Thus we conclude
that the "universal" distribution (\ref{PChapman}) can be derived
also within the general framework proposed in
Ref.~\cite{Aringazin4} that reflects a "universal" character of
the underlying $\chi^2$ distribution~\cite{Beck3}.

We note that although successful in describing the observed
statistics of Lagrangian acceleration, with a few simple
hypotheses and one fitting parameter, the one-dimensional Langevin
RIN models (\ref{P})-(\ref{FPconstant}) and
(\ref{Pnormal})-(\ref{LangevinExp}) suffer from the lack of
physical interpretation in the context of three-dimensional
Navier-Stokes equation.

In summary, we have presented a class of models
(\ref{Pnormal})-(\ref{LangevinExp}) using the basic assumption
that the parameter $\beta$ depends on normally distributed
velocity fluctuations. This class has been found to incorporate
the previous RIN models in a unified way, with the dependence
$\beta(u)$ required to be a (monotonic) Borel function of the
stochastic variable $u$.

\subsection{Hnat-Chapman-Rowlands model}\label{Sec:HCRmodel}

Another interesting model developed recently by Hnat, Chapman, and
Rowlands~\cite{Hnat} to describe the observed time series of the
solar wind bulk plasma parameters is based on the construction of
Fokker-Planck equation for which the probability density function
obeys the following one-parametric model-independent rescaling,
\be\label{Pscaling}
P(x,t) = t^{-\alpha_0}P_s(xt^{-\alpha_0}),
\ee
where $x$ denotes fluctuating plasma parameter and $\alpha_0$ is
the scaling index. The value $\alpha_0=1/2$ corresponds to a
self-similar Brownian walk with Gaussian probability density
functions at all time scales. The fitted value $\alpha_0=0.41$
corresponds to a single non-Gaussian distribution $P_s(x_s)$, to
which the observed distributions of some four plasma parameters
collapse under the scaling, $x_s=xt^{-\alpha_0}$.

The Langevin equation of this model assumes only additive noise,
and in such an anzatz it was found to be
\be
\partial_t x = D_1(x)+D_2(x)\eta(t),
\ee
where $D_1$ and $D_2$ are of the form
\be
D_1(x) = \sqrt{\frac{b_0}{D_0}}x^{1-\alpha_0^{-1}/2},
\ee
\be
D_2(x) = (b_0(1-\frac{1}{2}\alpha_0^{-1})-a_0)x^{1-\alpha_0^{-1}},
\ee
$a_0$, $b_0$ are constants, and $2D_0$ is intensity of the
delta-correlated Gaussian-white additive noise $\eta(t)$,
$\langle\eta(t)\rangle=0$. By construction, this specified form of
the dynamical equation ensures that the corresponding
Fokker-Planck equation,
\be\label{FPHnat}
\partial_t P(x,t)
 = \partial_x[a_0 x^{1-\alpha_0^{-1}}P + b_0x^{2-\alpha_0^{-1}}\partial_xP],
\ee
has the general solution $P(x,t)$, which exhibits the scaling
(\ref{Pscaling}). The fitted values are $a_0/b_0=2$, $b_0=10$, and
$\alpha_0^{-1}=2.44$. The rescaled distribution $P_s(x_s)$
corresponding to Eq.~(\ref{FPHnat}) is characterized by power law
tails truncated by stretched exponential with a good fit to the
tails of the experimental distribution, but it {\em diverges} at
the origin, $x_s\to 0$.

To sum up, we point out that this diffusion model uses the
generalized self-similarity principle resembling that used in the
Eulerian description of the energy cascade in developed
three-dimensional fluid turbulence, and appears to be valid only
asymptotically for large values of the variable, with the fitted
parameter value being about $\alpha_0=0.41$.


\subsection{Laval-Dubrulle-Nazarenko model}\label{Sec:LDNmodel}

The Navier-Stokes equation based approach to describe statistical
properties of small scale velocity increments, both in the
Eulerian and Lagrangian framework, was developed in much detail by
Laval, Dubrulle, and Nazarenko~\cite{Laval}; see also recent
work~\cite{Laval2}. This approach is based on featuring nonlocal
interactions between well separated large and small scales,
elongated triads, and referred to as the Rapid Distortion Theory
(RDT) approach. Decomposition of velocities into large and small
scale parts was made by introducing a certain spatial filter  of a
cut off type. Within the framework of this approach, a
three-dimensional Langevin model of the developed turbulence was
proposed.

In its one-dimensional version, the toy model of the Lagrangian
turbulence naturally implies a nonlinear Langevin type equation
for a component of the small scale velocity increments in time (in
the zero time-scale limit it corresponds to the acceleration $a$
of fluid particle)~\cite{Laval},
\be\label{LangevinLaval}
\partial_t a = (\xi - \nu_{\mathrm t}k^2)a + \sigma_\perp.
\ee
This equation is a Lagrangian description in the scale space, in
the reference frame comoving with a wave number packet. Here,
\be\label{turbviscosity}
\nu_{\mathrm t} = \sqrt{\nu_0^2+ B^2a^2/k^2}
\ee
stands for the turbulent viscosity introduced to describe small
scale interactions, $\nu_0$ is kinematic viscosity, $B$ is
constant, $k$ is wave number ($\partial_tk = -k \xi$, $k(0)=k_0$,
to model the RDT stretching effect in one-dimensional case), $\xi$
and $\sigma_\perp$ are multiplicative and additive noises
associated to the velocity derivative tensor and forcing of small
scales by large scales (the energy transfer from large to small
scales), respectively.

We refer to the model (\ref{LangevinLaval}) as the one-dimensional
Laval-Dubrulle-Nazarenko (LDN) model of the developed Lagrangian
turbulence. This toy model can also be viewed as a passive scalar
in a compressible one-dimensional flow.

The noises, one-dimensional versions of which appear in
Eq.~(\ref{LangevinLaval}), are projections related to the
large-scale velocities, $U_i$, and small-scale velocities, $u_i$,
of the flow as follows \cite{Laval},
\be\label{xi}
\hat\xi = \nabla(\frac{2\vec{k}}{k^2}(\vec{k}\vec{U})-\vec{U}),
\ee
\be\label{sigmaperp}
\hat\sigma_\perp = \hat{\sigma} - \frac{\vec
k}{k^2}(\vec{k}\hat\sigma),
\ee
\be\label{sigmai}
\sigma_i=\partial_j(\overline{U_iU_j} -U_iU_j + \overline{u_jU_i}
-\overline{U_ju_i}),
\ee
where the hat denotes Gabor transformation \cite{Nazarenko}, and
the bar stands for the spatial cut off retaining the large scale
part. One can see that the noises are related to the velocity
fluctuations and the additive noise contains interaction terms
between the large and small scale dynamics.

This gives a support to the idea that the intermittency is caused
also by some nonlocal interactions within the inertial range and
not merely by small scales. We remark that one would also like to
know the role of the dissipative scale in this integrated picture.

Noisy character of the entities (\ref{xi}) and (\ref{sigmaperp})
may not be seen as a consequence of the Navier-Stokes equation,
which does not contain external random forces at the
characteristic time scale. In the RDT approach, $\xi$ and
$\sigma_\perp$ are treated as independent stochastic processes
entering the small scale dynamics (\ref{LangevinLaval}) owing to
the fact that the large-scale dynamics is weakly affected by small
scales (that corresponds to a direct energy cascade in the
three-dimensional flow) and thus can be viewed as a given noise.

The relations (\ref{xi}) and (\ref{sigmaperp}) can be used to
trace back the origin of the multiplicative and additive noises
entering various surrogate Langevin models of the developed
turbulence, and to provide important information on the dynamics
underlying the intermittency.

Statistical properties of all the components of the noises $\xi$
and $\sigma_\perp$ were studied numerically for decaying
turbulence and reveal rich and complex behavior, in the laboratory
frame.

As a first step in the one-dimensional case, these noises were
modelled in the Lagrangian frame by the coupled delta-correlated
Gaussian white noises~\cite{Laval},
\begin{eqnarray}\label{noises}
\langle\xi(t)\rangle=0, \
\langle\xi(t)\xi(t')\rangle = 2D\delta(t-t'), \nonumber \\
\langle\sigma_\perp(t)\rangle = 0, \
\langle\sigma_\perp(t)\sigma_\perp(t')\rangle = 2\alpha\delta(t-t'), \\
\langle\xi(t)\sigma_\perp(t')\rangle = 2\lambda\delta(t-t'), \nonumber
\end{eqnarray}
where $D$, $\alpha$, and $\lambda$ are parameters depending on
scale via $k_0$. Stationary solution of the Fokker-Planck equation
associated to Eq.~(\ref{LangevinLaval}) with the noises
(\ref{noises}),
\begin{eqnarray}\label{FP}
\partial_t P(a,t) = \partial_a(\nu_{\mathrm t}k^2P)
 + D\partial_a (a \partial_a a P)
 - \lambda \partial_a (a\partial_a P)\nonumber \\
 -\lambda \partial^2_a (a P)
 +\alpha \partial^2_a P,\
\end{eqnarray}
is given by \cite{Laval}
\be\label{StationaryLaval}
P(a) = C\exp\left[ \int_{0}^{a}dy \frac{-\nu_{\mathrm t}k^2y - Dy
+\lambda}{Dy^2 -2\lambda y +\alpha}\right],
\ee
where $C$ is a normalization constant and six parameters can be
used to fit the experimental data. This model specifies the
one-dimensional LDN model (\ref{LangevinLaval}), and we refer to
this model as the LDN model with delta-correlated noises (dLDN
model).

The Langevin equation containing both the delta-correlated
Gaussian white multiplicative and additive noises was studied in
detail by Nakao \cite{Nakao} (see also references therein) by
using the associated Fokker-Planck equation. The dLDN model
(\ref{LangevinLaval}) extends Nakao's set up by incorporating two
new features: (i) the nonlinearity controlled by $B$ in
Eq.~(\ref{turbviscosity}) and (ii) the coupling of the noises
controlled by $\lambda$ in Eq.~(\ref{noises}).

It is interesting to note that the RDT approach qualitatively
resembles the model studied by Kuramoto and Nakao \cite{Kuramoto},
a system of spatially distributed chaotic elements driven by a
field produced by nonlocal coupling, which is spatially long-wave
and temporally irregular. Such systems, in which the
multiplicative noise is the local Lyapunov exponent fluctuating
randomly due to the chaotic motion of the elements, show power-law
correlations, intermittency, and structure functions similar to
that of the developed fluid turbulence.

In the next Section, we make a comparison of the RIN model
(\ref{P})-(\ref{FPconstant}) with the Laval-Dubrulle-Nazarenko
model (\ref{LangevinLaval}), as well as its particular case, the
dLDN model (\ref{StationaryLaval}).

\section{Comparison of the simple RIN and LDN models}\label{Sec:Comparison}

\subsection{A qualitative comparison}\label{Sec:QualitativeComparison}

A direct comparison of the Langevin equations (\ref{Langevin}) and
(\ref{LangevinLaval}) of the two models suggests the following
evident identifications:
\be\label{Comparison}
 \gamma F(a) = (\xi - \nu_{\mathrm t}k^2)a, \quad
 \sigma L = \sigma_\perp.
\ee
Hence the additive noises can be made identical to each other by
putting $\sigma^2=\alpha$. Further, in the case of a linear drift
force, $F(a)=-a$, and constant viscosity, $\nu_{\mathrm t}=\nu_0$,
we can identify the remaining parameters, $\gamma=\nu_0k^2-\xi$,
so that we get
\be\label{beta}
\beta \equiv \gamma/\sigma^2=(\nu_0k^2-\xi)/\alpha.
\ee
This relation implies that the parameter $\beta$ can be viewed as
a stochastic variable with a nonzero mean due to the stochastic
nature of $\xi$ assumed in the LDN model. This is in agreement
with the simple RIN model, the defining feature of which is just
that the fluctuating part of $\beta$ follows some statistical
distribution.

In the dLDN model (\ref{LangevinLaval})-(\ref{StationaryLaval}),
both the additive and multiplicative noises are taken {\em
delta-correlated} due to Eq.~(\ref{noises}). This is in a sharp
contrast to the assumption that $\beta$ can be taken constant to
derive the stationary solution (\ref{PGauss}) which is in the
foundation of the simple RIN model. More precisely, the solution
in the form (\ref{PGauss}) can be obtained as the lowest order
approximation if $\beta$ is slow varying in time as compared to a
typical time scale associated to the additive noise $L(t)$ (the
adiabatic approximation). This suggests that the multiplicative
noise $\xi$ should be taken sufficiently slow varying stochastic
variable, to meet the anzatz used in the RIN model.

The detailed numerical analysis of the noises \cite{Laval} for the
turbulent flow at relatively low Reynolds numbers,
$57<R_\lambda<80$, shows that the autocorrelation of the
multiplicative noise $\xi$ decays much slower (by about one order
of magnitude) than that of the additive noise $\sigma_\perp$.
Hence a typical time scale at which $\xi$ varies, $\tau_\xi$, is
considerably bigger than that, $\tau_\sigma$, of $\sigma_\perp$.
Also, the cross-correlation between the two noises was found to be
rather weak, $\lambda\ll D$ and $\lambda\ll \alpha$, by about two
orders of magnitude in the longitudinal case, and $\lambda=0$ in
the transverse case. Alltogether this allows one to introduce the
time scale hierarchy, $\tau_\xi \gg \tau_\sigma$, and to decouple
the noises, {\em i.e.} to put $\lambda=0$, that justifies the
adiabatic approximation and the one-dimensional RIN model.


The presence of the longtime correlated amplitude,
$e^{\omega(t)}$, and the short-time correlated directional part,
$L(t)$, of the stochastic driving force in Langevin type equation
considered by Mordant {\it et al.} \cite{Mordant} also supports
the above adiabatic approximation (two well separated time scales
in the single additive stochastic force, in the Lagrangian
framework). As usual, the delta-correlated noise originates from
taking the limit of zero correlation time in a system with the
smallest finite noise correlation time.

On the contrary, in the dLDN model one assumes the approximation
of comparable time scales, $\tau_\xi \simeq \tau_\sigma$, and
retains the coupling parameter $\lambda$ relating small scale
stretching and vorticity and responsible for the skewness, which
is however quite small in homogeneous isotropic turbulent flows.

The use of the constant turbulent viscosity, $\nu_{\mathrm
t}=\nu_0$, makes a good approximation in describing intermittency
corrections since both the constant and turbulent viscosity were
found to produce corrections which are of the same level as the
DNS result \cite{Laval}. In the physical context, this means that
the small scale interactions are not of much important in the
dynamics underlying the intermittency. This justifies the use of
the approximation of linear forcing, $F(a)=-a$, in the simple RIN
model. We note that this is also in an agreement with both the
experimental results for the Lagrangian velocity autocorrelation
function by Mordant, Metz, Michel, and Pinton \cite{Pinton}, and
the recent experimental Eulerian results for the spatial velocity
increments by Renner, Peinke, and Friedrich \cite{Friedrich}.

Alternatively, one can consider a {\it more general} RIN model
characterized by the presence of delta-correlated Gaussian white
additive and {\em multiplicative} noises and fluctuating
intensities of both the noises. This will lead to the model
similar to the dLDN model (\ref{LangevinLaval}) in which the noise
intensities $D$ and $\alpha$, and the coupling parameter $\lambda$
are assumed to fluctuate at a large time scale.

In summary, we found that the one-dimensional RIN model
(\ref{P})-(\ref{FPconstant}) can be viewed as a particular case of
the one-dimensional LDN model (\ref{LangevinLaval}) of turbulence
which is based on the RDT approach by Laval, Dubrulle, and
Nazarenko~\cite{Laval}. It should be stressed that while both the
toy models assume introduction of some external statistics ---the
correlator of $L(t)$ and the distribution $f(\beta)$ in
Eq.~(\ref{P}), and the correlators of $\xi$ and $\sigma_\perp$ in
Eq.~(\ref{LangevinLaval})--- the LDN model is characterized by a
solid foundation and reveals a rich structure as compared to the
RIN model.

In the first approximation, {\it i.e.}, $\lambda=0$, $\nu_{\mathrm
t}=\nu_0$, and $\tau_\xi \gg \tau_\sigma$, the class of RIN models
is in a quite good qualitative correspondence with the LDN model
(\ref{LangevinLaval}) and differs from the specific dLDN model
(\ref{LangevinLaval})-(\ref{StationaryLaval}) by the only fact
that in the latter one assumes $\tau_\xi \simeq \tau_\sigma$ and
introduces a delta-correlated multiplicative noise. Hence the
different resulting probability density functions for the
acceleration of fluid particle in the developed turbulent flow,
Eqs.~(\ref{PAringazin})-(\ref{PBeck}) and (\ref{StationaryLaval}),
respectively.


\subsection{A quantitative comparison}\label{Sec:QuantitativeComparison}

With the above result of the qualitative comparison, we are led to
make a more detailed, quantitative comparison of the dLDN model
(\ref{LangevinLaval})-(\ref{StationaryLaval}) and the simple RIN
model (\ref{P})-(\ref{FPconstant}) with the underlying $\chi^2$ or
log-normal distribution of $\beta$, in order to determine which
approximation, $\tau_\xi \simeq \tau_\sigma$ or $\tau_\xi \gg
\tau_\sigma$, is better when used to describe the Lagrangian
statistical properties of the developed turbulent flow. We take
the recent high precision Lagrangian experimental data
\cite{Bodenschatz,Bodenschatz2} on statistics of fluid particle
acceleration in the developed turbulent flow as a testbed.
Actually we follow the remark made in Ref.~\cite{Laval} that the
delta approximation of $\xi$ is debatable and the performance of
such a model should be further examined in the future.

In Ref.~\cite{Laval}, explicit analytic evaluation of the
distribution (\ref{StationaryLaval}) is given for the particular
case $\nu_{\mathrm t} =\nu_0$, while the general case is treated
in terms of $d\ln P(a)/da$ when fitting to the numerical RDT data
($-4 \leq a \leq 4$). In order to make fits to the experimental
probability density function $P(a)$ and to the contribution to the
fourth order moment, $a^4P(a)$, covering wide range of the
normalized acceleration, $-60\leq a\leq 60$, one needs in an
analytic or numerical evaluation of the r.h.s. of
Eq.~(\ref{StationaryLaval}). To this end, we have calculated {\em
exactly} the integral appearing in the dLDN probability density
function (\ref{StationaryLaval}) (see Appendix A).

The $\chi^2$ and log-normal distribution-based probability density
functions (\ref{PAringazin}) and (\ref{PBeck}) are both
realizations of the RIN model, and contain one fitting parameter,
$a_c$ and $s$, respectively. The result of comparison of fitting
qualities of these functions~\cite{Aringazin3}, with $a_c=39.0$
and $s=3.0$, is that the probability density function
(\ref{PBeck}) provides a better fit to the experimental data
\cite{Bodenschatz2} on low-probability tails and the contribution
to the kurtosis summarizing the peakedness of distribution.
However, since the integral in Eq.~(\ref{PBeck}) can not be
evaluated analytically we will use the
distribution~(\ref{PAringazin}), which provides a better fit to
the central region, when dealing with analytic expressions.

The dLDN probability density function (\ref{StationaryLaval})
contains six parameters which can be used for a fitting, the
multiplicative noise intensity $D$, the additive noise intensity
$\alpha$, the coupling $\lambda$ between the multiplicative and
additive noises, the turbulent viscosity parameter $B$, the
parameter $\nu_0$, and the wave number parameter $k$.

{\em The parameter $k$}. For a fitting, we can put $k=1$ without
loss of generality since it can be absorbed by the redefinition of
the parameters $\nu_0$ and $B$,
\be
\nu_0k^2 \to \nu_0, \quad Bk \to B.
\ee

{\em The parameter $\alpha$}.  The structure of the r.h.s. of
Eq.~(\ref{StationaryLaval}) is such that only four parameters out
of five can be used for a fitting. For example, one can put
$\alpha=1$ without loss of generality by using the following
redefinitions,
\be\label{redef}
\nu_0/\alpha \to \nu_0,\ B/\alpha \to B,\ D/\alpha \to D,\
\lambda/\alpha \to \lambda.
\ee
Alternatively, one can put $D=1$ provided the redefinitions
\be\label{redefD}
\nu_0/D \to \nu_0,\ B/D \to B,\ \alpha/D \to \alpha,\
\lambda/D \to \lambda.
\ee

{\em The parameter $\lambda$}. Due to Eq.~(\ref{A5}) we have
$c=-i\sqrt{D\alpha-\lambda^2}$, which is pure imaginary for
$D\alpha>\lambda^2$ and real for $D\alpha<\lambda^2$. For $c=0$,
{\it i.e.} $D\alpha=\lambda^2$, the integral
(\ref{LavalIntegral3}) is finite since divergent $F(c)$ and
$F(-c)$ defined by Eq.~(\ref{A4}) cancel each other. Since the
parameter $\lambda$ measuring the coupling between the noises is
assumed to be much smaller than both the noise intensities $D$ and
$\alpha$ \cite{Laval}, we put $D\alpha>\lambda^2$ in our
subsequent analysis. Moreover, the parameter $\lambda$ responsible
for the skewness can be set to zero since we will be interested,
as a first step, in statistically isotropic and homogeneous
turbulent flows, for which the experimental distribution $P(a)$
exhibits very small skewness~\cite{Bodenschatz}.

Thus, we can use three redefined free parameters, $\nu_0$, $B$,
and $D$, for a fitting, with $k=1$, $\alpha=1$, and $\lambda=0$.
However, we shall keep $k$ and $\alpha$ in an explicit way in the
formulas below, to provide a general representation.

We start by considering two important particular cases of the dLDN
probability density function (\ref{StationaryLaval}): a constant
viscosity, $\nu_{\mathrm t} = \nu_0$, and a dominating turbulent
viscosity, $\nu_{\mathrm t} = B|a|/k$.

\subsubsection{Constant viscosity}\label{Sec:Constant}

At $\lambda=0$ (symmetric case) and $B=0$, {\it i.e.} constant
viscosity $\nu_{\mathrm t} = \nu_0$, using
Eq.~(\ref{LavalIntegral1}) in Eq.~(\ref{StationaryLaval}) we get
(cf.~\cite{Laval})
\be\label{LavalPowerLaw}
P(a) = C(Da^2+\alpha)^{-(1+\nu_0k^2/D)/2},
\ee
where $C$ is normalization constant. This distribution is of a
power law type and we can compare it with the result
(\ref{PAringazin}), which contains a Gaussian truncation of
similar power law tails.

We note that with the identifications,
\be
 D/\alpha = 2(q-1), \quad
(1 + \nu_0k^2/D)/2 = 1/(q-1),
\ee
the distribution (\ref{LavalPowerLaw}) coincides with that
obtained in the context of generalized statistics with the
underlying $\chi^2$ distribution~\cite{Beck}. Particularly, for
$q=3/2$ ($n=3$, $\beta_0=3$) used there, it follows that
$D/\alpha=1$ and $\nu_0k^2=3$.

It is highly remarkable to note that the two different approaches
yield stationary distribution of exactly the same power law form,
for certain identification of the parameters. Namely, the Gaussian
white delta-correlated multiplicative and additive noises with
constant intensities and a linear drift term imply $P(a)$ of the
{same} form as that obtained in the RIN model with $\chi^2$
distributed $\beta$, the ratio of the drift coefficient to the
intensity of the Gaussian white delta-correlated additive noise.
It follows that the effect of $\chi^2$ distributed $\beta$ mimics
the presence of the multiplicative noise, and {\it vice versa}, in
this particular case.

The power law distribution (\ref{LavalPowerLaw}) can be used to
get a good fit of the Lagrangian experimental $P(a)$ data for
small accelerations, {\em e.g.}, with the normalized values
ranging from $-10$ to $10$, but in contrast to the Gaussian
truncated one (\ref{PAringazin}) it exhibits strong deviations for
large $a$, and for $(1 + \nu_0k^2/D)/2\leq 2$ leads to a divergent
fourth order moment, which is known to be finite
\cite{Bodenschatz,Aringazin2,Aringazin3}.

Introducing the noise intensity ratio parameter,
\be
b=\sqrt{D/\alpha},
\ee
and denoting
\be
\kappa = -(1 + \nu_0k^2/D)/2
\ee
we can rewrite the normalized distribution (\ref{LavalPowerLaw})
as follows (cf.~\cite{Nakao})
\be\label{PNakao}
P(a) = \frac{(1+b^2a^2)^\kappa}
            {2 {_2\!}F_1(-\kappa;\frac{1}{2};\frac{3}{2};-b^2)},
\ee
where ${_2\!}F_1$ is the hypergeometric function. In accord to the
analysis made by Nakao~\cite{Nakao}, for {\em small} additive
noise intensity, {\it i.e.}, at $b \gg 1$, this distribution
exhibits a pronounced plateau near the origin, and the $n$th order
moments, {\em truncated} by reflective walls at some fixed $|a|$,
behave as a power of $b$,
\begin{eqnarray}
\langle a^n\rangle \sim b^{-\nu_0k^2/D}, {\textrm{ for } }
n>\nu_0k^2/D,\\
\langle a^n\rangle \sim b^{-n}, {\textrm{ for } }
n<\nu_0k^2/D,
\end{eqnarray}
where $n>0$. Thus, the truncated moments behave as
\be\label{anNakao}
\langle a^n\rangle \sim G_0 + G_1b^{-H(n)},
\ee
where $G_{0,1}$ are some constants and the function $H(n)$ is zero
at $\nu_0k^2=0$ and monotonically saturates to $n$ at big
$\nu_0k^2$. It should be stressed that such a behavior of the
moments for small additive noise intensity is not specific to the
distribution (\ref{PNakao}) since it gives divergent moments but
arises after some truncation of it, for example, by means of
reflective walls or nonlinearity. Particularly, a truncation of
the power law tails of the distribution naturally arises when
accounting for the turbulent viscosity to which we turn below.

\subsubsection{Dominating turbulent viscosity}\label{Sec:Dominating}

At $\lambda=0$ (symmetric case), for the case of dominating
turbulent viscosity, $\nu_{\mathrm t} = B|a|/k$, using
Eq.~(\ref{LavalIntegral2}) we get for positive and negative $a$,
respectively,
\be\label{LavalExpLaw}
P(a) = \frac{Ce^{\mp Bka/D \pm {Bk\alpha^{1/2}}{D^{-3/2}}
 \mathrm{arctan}[(D/\alpha)^{1/2}a]}}
 {(Da^2+\alpha)^{1/2}},
\ee
where $C$ is normalization constant. One can see that, as
expected, the power law dependence is of a similar form as in
Eq.~(\ref{LavalPowerLaw}) but it is exponentially truncated at big
$|a|$ owing to the turbulent viscosity term. This distribution is
similar to the Gaussian truncated one (\ref{PAringazin}) but the
truncation is of an exponential type and there is some symmetric
enhancement of the tails supplied by the $\mathrm{arctan}$ term.

Now we turn to the general case, which provides a link between the
two particular cases, $\nu_{\mathrm t} = \nu_0$ and $\nu_{\mathrm
t} = B|a|/k$, considered above.

\subsubsection{The general symmetric case}\label{Sec:GeneralCase}

At $\lambda=0$ (symmetric case), from  Eqs.~(\ref{A5})-(\ref{A7})
we have
\be\label{c12}
 c=-id_2,\
 c_1=id_1^2d_2,\
 c_2=kd_1,\
\ee
where we have denoted
\be\label{d12}
 d_1=\sqrt{D(Dk^2\nu_0^2-B^2\alpha)},\quad
 d_2=\sqrt{D\alpha}.
\ee
Note that $c$ is pure imaginary and the r.h.s. of
(\ref{LavalIntegral3}) is much simplified yielding a symmetric
distribution with respect to $a\to-a$. The entity $c_2$ defined by
Eq.~(\ref{A7}) may be either real (for $Dk^2\nu_0^2>B^2\alpha$) or
pure imaginary (for $Dk^2\nu_0^2<B^2\alpha$). In particular, for
the case of constant viscosity, $\nu_0^2 \gg B^2$, it is real
while for the case of dominating turbulent part of the viscosity,
$B^2 \gg \nu_0^2$, it is pure imaginary, provided that the
intensities of noises, $D$ and $\alpha$, are of the same order of
magnitude. These two particular cases lead to different final
expressions for the distribution, (\ref{LavalPowerLaw}) and
(\ref{LavalExpLaw}), respectively, obtained above.

In the general case, using Eqs.~(\ref{c12}) and (\ref{d12}) in
Eq.~(\ref{LavalIntegral3}) after some algebra we obtain the
following expression for the dLDN probability density function
(\ref{StationaryLaval}), at $\lambda=0$,
\begin{eqnarray}\label{LavalGenLaw}
P(a) = \frac{Ce^{-\nu_{\mathrm t}k^2/D}}
 {(Da^2+\alpha)^{1/2}}\nonumber \\
 \times\left[
 \frac{4D^5(B^4D\alpha a^2 +k^2(Dk\nu_0^2 +d_1\nu_{\mathrm t})^2)}
 {d_1^6k^2(Da^2+\alpha)}
 \right]^{kd_1/(2D^2)},
\end{eqnarray}
where $C$ is normalization constant, $d_1$ is given in
Eq.~(\ref{d12}), and $\nu_{\mathrm t}$ is given by
Eq.~(\ref{turbviscosity}).

It can be easily checked that Eq.~(\ref{LavalGenLaw}) reduces to
Eq.~(\ref{LavalPowerLaw}) at $B=0$, while to verify that it
reduces to Eq.~(\ref{LavalExpLaw}) at $\nu_{\mathrm t} = B|a|/k$
requires the use of the fact that $d_1$ becomes pure imaginary,
returning back to the logarithmic representation due to
Eq.~(\ref{A4}), and the identity~(\ref{A8}).

The distribution (\ref{LavalGenLaw}) is characterized by the power
law tails, which are (i) exponentially truncated and (ii) enhanced
by the power law part of the numerator, with both the effects
being solely related to the nonzero turbulent viscosity
coefficient $B$ responsible for a nonlinear small scale dynamics.

We conclude that to provide an acceptable fit of the dLDN model
prediction to the Lagrangian experimental data \cite{Bodenschatz2}
{\em nonlinear} small scale interactions encoded in the turbulent
viscosity $\nu_{\mathrm t}$ are essential.

\begin{figure}[tbp!]
\begin{center}
\includegraphics[width=0.45\textwidth]{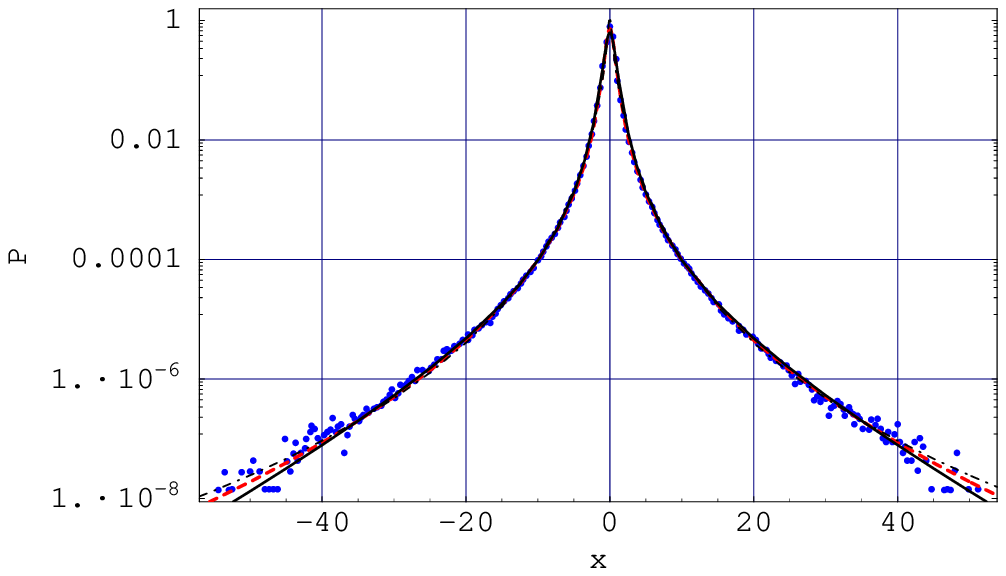}
\includegraphics[width=0.45\textwidth]{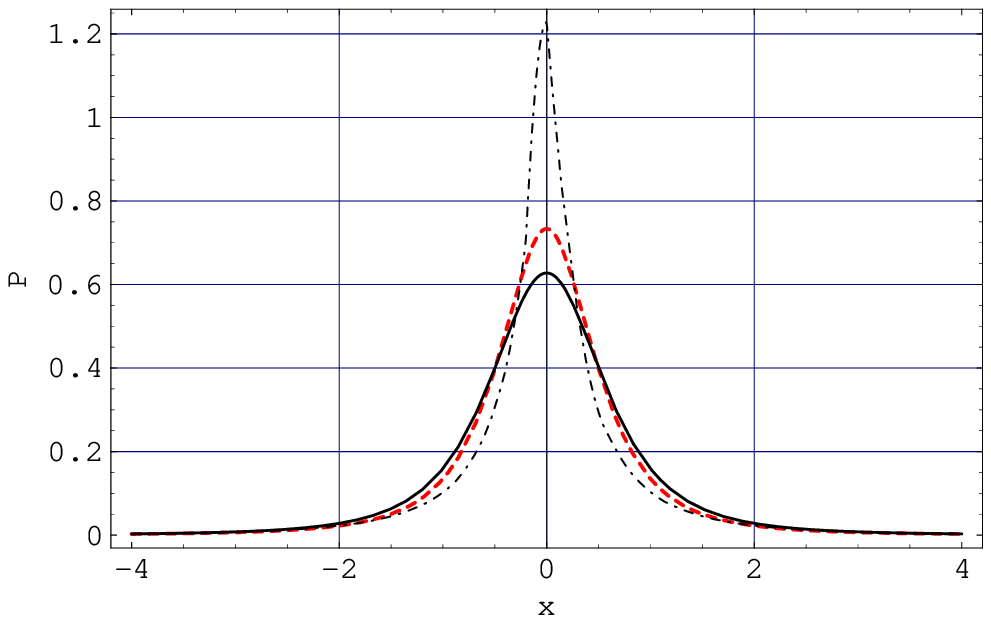}
\end{center}
\caption{ \label{Fig5} Acceleration probability density function
$P(a)$. Dots: experimental data at $R_\lambda=690$ by Crawford,
Mordant, and Bodenschatz~\cite{Bodenschatz2}.  Dashed line:
stretched exponential fit (\ref{Pexper}), $b_1=0.513$, $b_2=
0.563$, $b_3= 1.600$, $C=0.733$. Dot-dashed line: Beck log-normal
model (\ref{PBeck}), $s=3.0$. Solid line: Laval-Dubrulle-Nazarenko
model (\ref{LavalGenLaw}), $k = 1$, $\alpha=1$, $D = 1.130$, $B =
0.163$, $\nu_0 = 2.631$, $C=1.805$. $x=a/\langle a^2
\rangle^{1/2}$ denotes normalized acceleration.}
\end{figure}

\begin{figure}[tbp!]
\begin{center}
\includegraphics[width=0.45\textwidth]{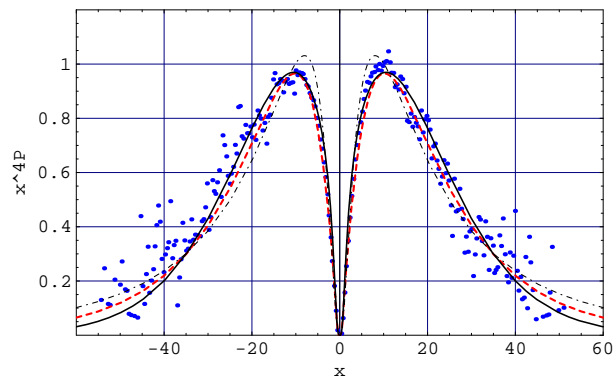}
\end{center}
\caption{ \label{Fig6} Contribution to fourth order moment
$a^4P(a)$. Notation is the same as in Fig.~\ref{Fig5}.}
\end{figure}

Sample fit of the dLDN probability density function $P(a)$ given
by Eq.~(\ref{LavalGenLaw}) and contribution to the fourth order
moment, $a^4P(a)$, are shown in Figs.~\ref{Fig5} and \ref{Fig6},
respectively. In the numerical fit, we have put, in accord to the
redefinitions (\ref{redef}), the wave number parameter $k=1$ and
the additive noise intensity parameter $\alpha=1$ in
Eq.~(\ref{LavalGenLaw}) and fitted the remaining three parameters,
$\nu_0$, $D$, and $B$. One can observe a good agreement with the
experimental data. Particularly, the dLDN contribution to the
kurtosis $a^4P(a)$ plotted in Fig.~\ref{Fig6} does peak at the
same points as the experimental curve (positions of the peaks
depend mainly on $D$). The central part of the dLDN distribution
shown in the bottom panel of Fig.~\ref{Fig5} fits the experiment
to a higher accuracy as compared with the log-normal model
(\ref{PBeck}) but yet departures from that of the experimental
curve. This departure can be attributed to the approximation of
delta-correlated multiplicative noise used in the dLDN model (see
discussion in Sec.~\ref{Sec:Comparison} above).

Having the general form of the dLDN distribution evaluated
explicitly, Eq.~(\ref{LavalGenLaw}), one can derive higher
acceleration moments, $\langle a^{n}\rangle$, $n=2,4,\dots$ The
associated integrals are not analytically tractable and can be
evaluated numerically. We will consider these in
Sec.~\ref{Sec:ConditionalProbability} below.

In the most general case ($\lambda\not=0$) the resulting $P(a)$ is
given due to an exponential of the exact integral
(\ref{LavalIntegral3}) which we do not represent here for brevity.

To sum up, we have made an important step forward with the dLDN
model by having calculated $P(a)$ exactly. We have shown that the
dLDN model is capable to reproduce the recent Lagrangian
experimental data on the acceleration statistics to a good
accuracy. Particularly, we found that the predicted fourth order
moment density function does peak at the same value of
acceleration, $|a|/\langle a^2\rangle^{1/2}\simeq 10.2$, as the
experimental curve, in contrast to the predictions of the other
considered stochastic models. The presence of the delta-correlated
multiplicative noise and the nonlinearity (turbulent viscosity) in
the model Langevin equation was found to be of much importance.
The considered RIN models provide less but yet acceptable accuracy
of the low-probability tails despite they employ only one free
parameter, which can be fixed by certain phenomenological
arguments, as compared to the dLDN model, which contains four free
parameters. However, we stress that in contrast to the LDN model
the considered RIN models have a meager support from the
turbulence dynamics.

\section{Conditional probability density function of the acceleration}
\label{Sec:ConditionalProbability}

In the very recent paper \cite{Mordant0303003} a new experimental
data on the conditional probability density function of the
transverse acceleration, $P(a|u)$, have been reported. In general,
this representation is in agreement with the proposed idea that
velocity fluctuations $u$ are directly involved in the stochastic
acceleration dynamics \cite{Aringazin4} represented in
Sec.~\ref{Sec:GaussianDistribution}.

Also, the observed conditional acceleration component variance has
been found to be in a good agreement with the Sawford {\it et al.}
scaling relation (see \cite{Mordant0303003} and references
therein),
\be\label{a2u}
\langle a^2|u\rangle \sim u^6,
\ee
obtained to a leading order in the same component $u$ (to be not
confused with the {\it rms} velocity $\bar u=\langle
u^2\rangle^{1/2}$).

The experimental data reveal highly non-Gaussian, stretched
exponential character of $P(a|u)$, very similar to that of $P(a)$,
for fixed $u$ ranging from zero up to three {\it rms} velocity
$\bar u$~\cite{Mordant0303003} as opposed to the theoretical
result that $P(a|u)$ is a Gaussian in $a$ due to the simple RIN
model~(\ref{PGauss}), with arbitrary $\beta=\beta(u)$, or due to
the more general RIN model~(\ref{gu}). Similarity between the
experimental $P(a|u)$ and $P(a)$ suggests that they share the
process underlying the fluctuations.

Below we address this important problem within the framework of
the RIN approach.

The idea is that the stretched exponential form of the tails of
the observed conditional distribution $P(a|u)$ could be assigned
solely to small time scales, while the marginal probability
distribution $P(a)$ is developed from $P(a|u)$ at large time
scales, in accord to the two time-scale dynamics.

This requires some modification in the simple RIN models.  The
sole use of the delta-correlated Gaussian white additive noise,
with fluctuating intensity depending on $u$, and a linear force,
$F(a)=-a$, with fluctuating $\gamma=\gamma(u)$, is not capable to
explain the stretching in the observed $P(a|u)$, as it implies
only Gaussian conditional probability density function $P(a|u)$,
for any fixed $u$.

However, it is known that accounting for the {\em multiplicative}
delta-correlated Gaussian white noise in the drift term of
Langevin equation implies stretched exponential tails.

Hence we can simply follow the dLDN anzatz as a constitutive model
(see Sec.~\ref{Sec:LDNmodel}) using the assumption that the
additive noise intensity $\alpha$ appearing in the stationary
probability distribution $P(a|D,\alpha,B,\nu_0)$ given by
Eq.~(\ref{LavalGenLaw}) depends on $u$.

Here, we put the parameter $\lambda$ measuring coupling of the
noises to each other to zero ignoring thus the skewness effect
which is very small for both the experimental $P(a|u)$ and
$P(a)$~\cite{Bodenschatz,Mordant0303003}. This effect is
nevertheless of much interest since it is associated to the
relationship between stretching and vorticity in a
three-dimensional flow. More important here is that it may imply
additional stretching of the tails as well (see $\lambda$
dependent terms in Eq.~(\ref{LavalIntegral3})). This possible way
to explain stretched exponential tails of the observed $P(a|u)$
can be considered elsewhere.

Following the arguments and techniques presented in
Sec.~\ref{Sec:GaussianDistribution}, the marginal probability
distribution $P(a)=P(a|D,B,\nu_0)$ is obtained by integrating out
$u$ in
\be\label{Pu}
P(a|u)=P(a|D,\alpha(u),B,\nu_0),
\ee
with an appropriate choice of the function $\alpha(u)$, for
example, $\alpha(u)=e^{u}$, and some probability distribution of
$u$, for example, a Gaussian one with zero mean. Note that once
$P(a|D,\alpha(u),B,\nu_0)$ is fitted to the experimental curves of
the conditional $P(a|u)$ there formally remains only one parameter
to be fitted in the marginal $P(a)$, the variance of the Gaussian
distribution of $u$, {\it i.e.}, the {\it rms} velocity $\bar u$.

\begin{figure}[tbp!]
\begin{center}
\includegraphics[width=0.45\textwidth]{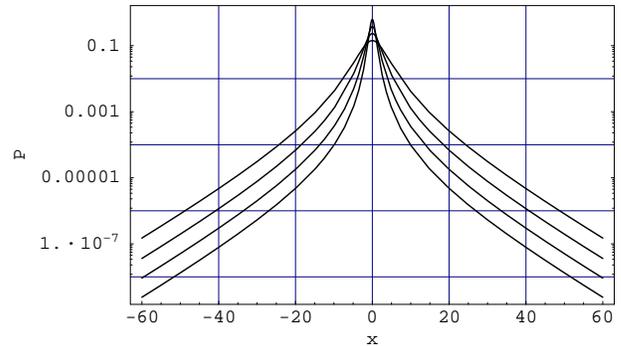}
\end{center}
\caption{ \label{Fig7} Conditional probability density function
$P(a|u)$ given by Eq.~(\ref{LavalGenLaw}) for $\alpha=e^u$, $k =
1$, $D = 1.130$, $B = 0.163$, $\nu_0 = 2.631$. The inner curve:
$u=0$. The outer curve: $u=3$. $x=a/\langle a^2 \rangle^{1/2}$
denotes normalized acceleration.}
\end{figure}

The normalized conditional distribution (\ref{Pu}) given by
Eq.~(\ref{LavalGenLaw}) with $\alpha=e^u$ is shown in
Fig.~\ref{Fig7}, for four values, $u=0,1,2,3$, and the other
parameters fixed. One can observe an increase of the variance with
the increase of $u$, and a good qualitative agreement with the
experimental curves $P(a|u)$~\cite{Mordant0303003}. We note that
the velocity fluctuations are present only in the definition of
the LDN additive noise (\ref{sigmai}) which can be viewed as a
hint that only additive noise intensity essentially depends on
$u$. These results give an independent support to the model
represented in Sec.~\ref{Sec:GaussianDistribution}.

\begin{figure}[tbp!]
\begin{center}
\includegraphics[width=0.45\textwidth]{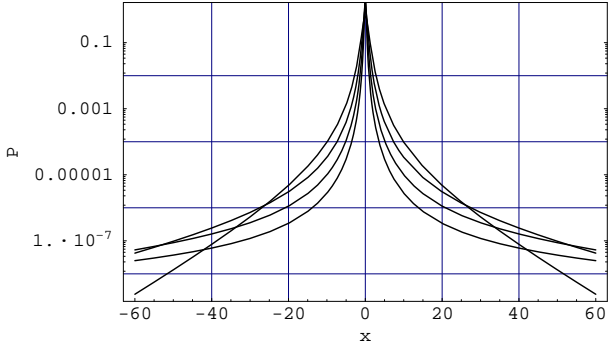}
\end{center}
\caption{ \label{Fig8} Conditional probability density function
$P(a|u)$ given by Eq.~(\ref{LavalGenLaw}) for $\alpha=1$, $k = 1$,
$D = 1.13 e^u$, $B = 0.163$, $\nu_0 = 2.631$. The outer curve:
$u=0$. The inner curve: $u=3$. $x=a/\langle a^2 \rangle^{1/2}$.}
\end{figure}

\begin{figure}[tbp!]
\begin{center}
\includegraphics[width=0.45\textwidth]{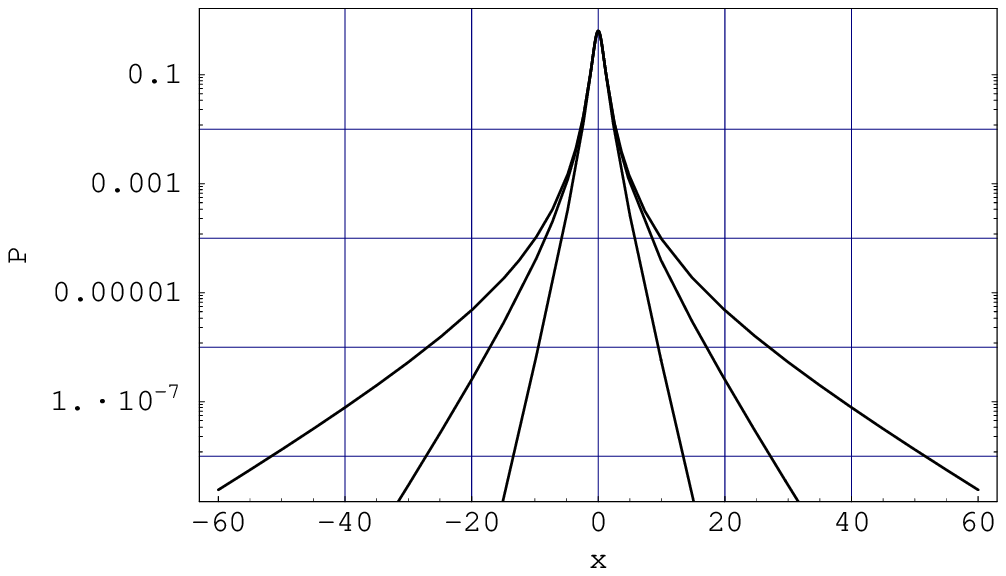}
\includegraphics[width=0.45\textwidth]{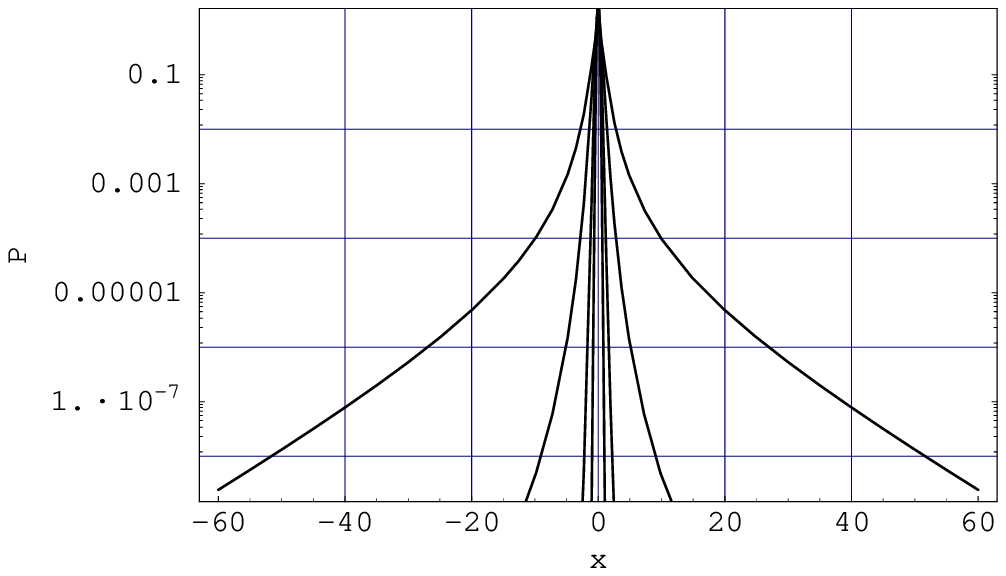}
\end{center}
\caption{ \label{Fig9} Conditional probability density function
$P(a|u)$ given by Eq.~(\ref{LavalGenLaw}). Top panel: $\alpha=1$,
$k = 1$, $D = 1.130$, $B = 0.163e^{u}$, $\nu_0 = 2.631$ (the outer
curve: $u=0$, the inner curve: $u=2$). Bottom panel: $\alpha=1$,
$k = 1$, $D = 1.130$, $B = 0.163$, $\nu_0 = 2.631e^{u}$ (the outer
curve: $u=0$, the inner curve: $u=3$). $x=a/\langle a^2
\rangle^{1/2}$.}
\end{figure}

We have checked a different reasonable assumption that the
multiplicative noise intensity parameter $D$ depends on the
velocity fluctuations, $D\simeq e^u$, with the other parameters
fixed. The result is shown in Fig.~\ref{Fig8}. One can observe
that the change of the shape of $P(a|u)$ defined as
$P(a|D(u),\alpha,B,\nu_0)$ with the increase of $u$ does not
qualitatively meet that observed in the
experiments~\cite{Mordant0303003}. We note that an increase of
$D$, {\it i.e.}, stronger multiplicative noise, is generally
understood as the pronounced increase of the {relative} chance for
a fluid particle to have higher accelerations as compared to low
accelerations, due to the multiplicative random process. This
point of view is confirmed by Fig.~\ref{Fig8}. Also, for a
completeness in Fig.~\ref{Fig9} we represent sample dependencies
of $P(a|D,\alpha,B,\nu_0)$ on the parameters $B$ and $\nu_0$.

\begin{figure}[tbp!]
\begin{center}
\includegraphics[width=0.45\textwidth]{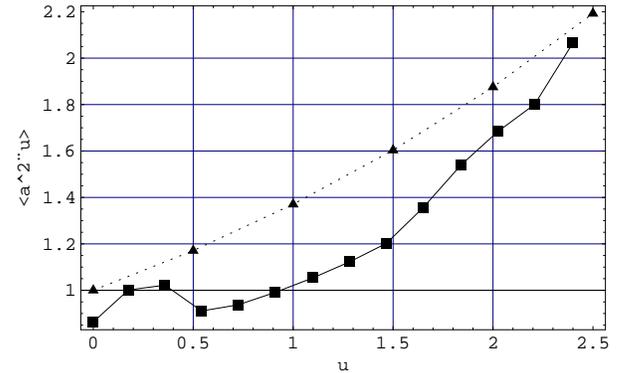}
\end{center}
\caption{ \label{Fig10} The normalized conditional acceleration
variance, $\langle a^2|u\rangle$ as a function of normalized
velocity fluctuations $u/\langle u\rangle^{1/2}$. Boxes: the
experimental data on $\langle a^2|u\rangle/\langle
a^2\rangle$~\cite{Mordant0303003}, triangles: $\langle
a^2|u\rangle/\langle a^2|0\rangle$ with $P(a|u)$ given by
Eq.~(\ref{LavalGenLaw}) for $\alpha=e^{u/3}$, $k = 1$, $D =
1.130$, $B = 0.163$, $\nu_0 = 2.631$.}
\end{figure}

Using the dLDN probability function $P(a|D,\alpha(u),B,\nu_0)$ one
can compute $\langle a^2|u\rangle$ and compare the result with the
predicted scaling relation (\ref{a2u}) and the experimental
data~\cite{Mordant0303003}. A sample plot of $\langle
a^2|u\rangle$ is shown in Fig.~\ref{Fig10}, where we have used the
exponential in the form $\alpha=e^{u/u_0}$ to provide possibility
for a fitting to the experimental data (boxes). Although with the
fitted exponent, $u_0=3$, the result (triangles) exhibits a
departure to the experiment, qualitatively the model implies a
correct behavior of the conditional acceleration variance.

To summarize, the observed stretched exponential form of the
conditional acceleration probability density function $P(a|u)$ can
be understood within the framework of the dLDN
model~(\ref{LavalGenLaw}) due to the effect of the multiplicative
noise under the assumption that the additive noise intensity
$\alpha$ depends on velocity fluctuations $u$. The alternative
assumption that the multiplicative noise intensity depends on $u$
seems not to be in a qualitative agreement with the shapes of
experimental curves at the different values of $u$ except for
$u=0$. The predicted conditional acceleration variance $\langle
a^2|u\rangle$ with $\alpha=e^{u/u_0}$, $u_0=3$, have been found in
a good qualitative agreement with the experimental curve. However,
we observe a departure to the experimental data.

\section{Discussion}

We conclude with a few remarks.

(i) It is interesting to note that a "universal" probability
density function $f(u)$ could be identified with the help of the
relation,
\be\label{u92}
\int_{-\infty}^{\infty} \!\!\!\! du\ u^6 f(u) \sim {\bar u}^{9/2},
\ee
stemming from a comparison of the Heisenberg-Yaglom scaling
relation (\ref{HYscaling}) and the scaling (\ref{a2u}), both
recently confirmed by the experiments, where we assume that
$\int_{-\infty}^{\infty} du\,
\langle a^2|u\rangle f(u) \sim \langle a^2\rangle$. Obviously, the
choice of a normalized Gaussian probability density function with
zero mean for $f(u)$ does not lead to the above relation since it
implies $\sim {\bar u}^6$. This can be viewed as a signal for
selecting a different probability distribution of $u$, such as
that of a stretched exponential form.

(ii) In the present paper, we put the parameter $\lambda$, which
measures the coupling between the multiplicative and additive
noises in the dLDN model, to zero discarding thus skewness effects
in the predicted $P(a|u)$ and $P(a)$. This effect is of much
interest to study in order to estimate $\lambda$ using the fitting
to Lagrangian experimental data on the longitudinal component of
acceleration with respect to the trajectory ($\lambda=0$ by
definition for the transverse component of acceleration).
Presumably it is small due to small skewness of both the observed
$P(a|u)$ and $P(a)$ for the measured $x$ component of
acceleration~\cite{Bodenschatz2}.

(iii) Following the RIN approach presented in this paper it is of
interest to evaluate the proposed averaging of the dLDN
probability distribution $P(a|D,\alpha(u),B,\nu_0)$ given by
Eq.~(\ref{LavalGenLaw}) over normally distributed $u$ with
$\alpha$ taken to be $\alpha=e^u$. This is equivalent to the
averaging over log-normally distributed $\alpha$. A comparison of
the resulting distribution with the Lagrangian experimental data
can be made elsewhere.

(iv) In the present paper we have not reviewed a recent work by
Reynolds~\cite{Reynolds}. A comparison of the results of the
Reynolds model with that of the stochastic models proposed in
\cite{Beck,Beck4} can be found in the recent paper by Mordant {\it
et al.}~\cite{Mordant0303003}.

\appendix

\section{Exact integrals}\label{Sec:ExactIntegrals}

Exact indefinite integrals, up to a constant term which does not
depend on $a$, used in calculating the definite integral entering
the probability density function (\ref{StationaryLaval}) are given
below.

At $\nu_{\mathrm t} = \nu_0$,
\begin{eqnarray}\label{LavalIntegral1}
\int\!\!\! da \frac{-\nu_0k^2a - Da +\lambda}{Da^2 -2\lambda a
+\alpha}
 = -\frac{D+\nu_0k^2}{2D} \ln[Da^2\!-\!2\lambda a \!+\!\alpha] \nonumber
 \\
 + \frac{\lambda\nu_0k^2}{D\sqrt{D\alpha-\lambda^2}}\
 \mathrm{arctan}\frac{Da-\lambda}{\sqrt{D\alpha-\lambda^2}}.\quad
\end{eqnarray}

At $\nu_{\mathrm t} = B|a|/k$, for positive and negative $a$,
respectively,
\begin{eqnarray}\label{LavalIntegral2}
\int da \frac{\mp Bka^2 - Da +\lambda}{Da^2 -2\lambda a +\alpha}
 = \mp\frac{B k a}{D} \nonumber \\
 -\frac{D^2 \pm 2B\lambda k}{2D^2} \ln[Da^2-2\lambda a +\alpha]
 \nonumber \\
 \pm \frac{B(D\alpha-2\lambda^2)k}{D^2\sqrt{D\alpha-\lambda^2}}\
 \mathrm{arctan}\frac{Da-\lambda}{\sqrt{D\alpha-\lambda^2}},
\end{eqnarray}

In the general case, we have obtained a cumbersome expression,
\begin{eqnarray}\label{LavalIntegral3}
\int \!\! da \frac{-\nu_{\mathrm t}k^2a - Da +\lambda}{Da^2
-2\lambda a +\alpha}
 = -\frac{\nu_{\mathrm t}k^2}{D}
   -\frac{1}{2}\ln[Da^2\!-\!2\lambda a \!+\!\alpha] \nonumber \\
   -\frac{2B\lambda k}{D^2}\ln[2Bka+\nu_{\mathrm t}k^2]
   +F(c) + F(-c), \quad
\end{eqnarray}
where we have denoted
\begin{eqnarray}\label{A4}
F(c)
 = \frac{c_1k^2}{2c_2D^2c}\ln[\frac{2D^3}{c_1c_2(c-Da+\lambda)}\times
   \nonumber \\
   \times(
   B^2(\lambda^2 + c\lambda-D\alpha)a
   + c(D\nu_{\mathrm t}^2k^2+c_2\nu_{\mathrm t})
    )
   ],
\end{eqnarray}
\be\label{A5}
c=-i\sqrt{D\alpha-\lambda^2}, \quad \nu_{\mathrm t} =
\sqrt{\nu_0^2+ B^2a^2/k^2},
\ee
\be\label{A6}
c_1 = B^2(4\lambda^3+4c\lambda^2 - 3D\alpha\lambda-cD\alpha)
    + D^2(c +\lambda)\nu_0^2k^2,
\ee
\be\label{A7}
c_2 = \sqrt{B^2(2\lambda^2 + 2c\lambda-D\alpha)k^2 +
D^2\nu_0^2k^4}.
\ee
Some useful formulas used in verifying the limits $B\to 0$ and
$D\to 0$ are:
\be\label{A8}
\mathrm{arctan}\, x = \frac{i}{2}(\ln(1-ix)- \ln(1+ix)),
\ee
\be\label{A9}
\lim_{D\to 0}\frac{1}{D}\ln[1+Da^2] = a^2.
\ee


\begin{thebibliography}{32}

\bibitem{Tsallis}
C. Tsallis, J. Stat. Phys. {\bf 52}, 479 (1988).

\bibitem{Johal}
R. Johal, {\it An interpretation of Tsallis statistics based on
polydispersity}, cond-mat/9909389 (1999).

\bibitem{Aringazin}
A.K. Aringazin and M.I. Mazhitov, Physica A {\bf 325}, 409 (2003);
{\it Quasicanonical Gibbs distribution and Tsallis nonextensive
statistics}, cond-mat/0204359 (2002).

\bibitem{Beck3}
C. Beck and E.G.D. Cohen, Physica A {\bf 322}, 267 (2003); {\it
Superstatistics}, cond-mat/0205097 (2002).

\bibitem{Beck4}
C. Beck, {\it Lagrangian acceleration statistics in turbulent
flow}, cond-mat/0212566 (2002).

\bibitem{Beck}
C. Beck, Phys. Rev. Lett. {\bf 87}, 180601 (2001); {\it
Generalized statistical mechanics and fully developed turbulence},
cond-mat/0110073 (2001).

\bibitem{Wilk}
G. Wilk and Z. Wlodarczyk, Phys. Rev. Lett. {\bf 84}, 2770 (2000).

\bibitem{Beck2}
 C. Beck, Physica  A {\bf 277}, 115 (2000);
 Phys. Lett. A {\bf 287}, 240 (2001);
 Europhys. Lett. {\bf 57}, 329 (2002);
 {\it Non-additivity of Tsallis entropies and fluctuations of
temperature}, cond-mat/0105371 (2001).

\bibitem {Reynolds}
A.M. Reynolds,
Phys. Fluids {\bf 15}, L1 (2003).

\bibitem{Aringazin2}
A.K. Aringazin and M.I. Mazhitov, {\it Phenomenological Gaussian
screening in the nonextensive statistics approach to fully
developed turbulence}, cond-mat/0212462 (2002).

\bibitem{Aringazin3}
A.K. Aringazin and M.I. Mazhitov, {\it Gaussian factor in the
distribution arising from the nonextensive statistics approach to
fully developed turbulence}, cond-mat/0301040 (2003).

\bibitem{Aringazin4}
A.K. Aringazin and M.I. Mazhitov, Phys. Lett. A {\bf 313}, 284
(2003); {\it The PDF of fluid particle acceleration in turbulent
flow with underlying normal distribution of velocity
fluctuations}, cond-mat/0301245 (2003).

\bibitem{Kraichnan0305040}
T. Gotoh and R.H. Kraichnan, {\it Turbulence and Tsallis
statistics}, nlin.CD/0305040 (2003).

\bibitem{Castaing}
B. Castaing, Y. Gagne, and E.J. Hopfinger, Physica D {\bf 46}, 177
(1990).

\bibitem{Sawford}
 B.L. Sawford, Phys. Fluids A {\bf 3}, 1577 (1991).
 S.B. Pope, Phys. Fluids {\bf 14}, 2360 (2002).

\bibitem{K62}
A.N. Kolmogorov, J. Fluid. Mech. {\bf 13}, 82 (1962).
L.D.~Landau and E.M.~Lifschitz, {\it Fluid mechanics}, 2nd Ed.
(Pergamon Press, Oxford, 1987).

\bibitem{Bodenschatz}
A. La Porta, G.A. Voth, A.M. Crawford, J. Alexander, and
E.~Bodenschatz, Nature {\bf 409}, 1017 (2001). G.A.~Voth, A.~
La~Porta, A.M.~Crawford, E.~Bodenschatz, J.~Alexander, J.~Fluid
Mech. {\bf 469}, 121 (2002); {\it Measurement of particle
accelerations in fully developed turbulence}, physics/0110027
(2001).

\bibitem{Bodenschatz2}
A.M. Crawford, N. Mordant, E. Bodenschatz, and A.M. Reynolds, {\it
Comment on "Dynamical foundations of nonextensive statistical
mechanics"}, physics/0212080 (2002), submitted to Phys. Rev. Lett.

\bibitem{Mordant}
N. Mordant, J. Delour, E. Leveque, A. Arneodo, and J.-F. Pinton,
Phys. Rev. Lett. {\bf 89}, 254502 (2002); {\it Long time
correlations in Lagrangian dynamics: a key to intermittency in
turbulence}, physics/0206013 (2002).

\bibitem{Mordant0303003}
N. Mordant, A.M. Crawford, and E. Bodenschatz, {\it Experimental
Lagrangian acceleration probability density function measurement},
physics/0303003 (2003).

\bibitem{Gotoh}
R.H. Kraichnan and T. Gotoh, data presented at the CNLS Workshop
on {\it Anomalous Distributions, Nonlinear Dynamics, and
Nonextensivity}, Santa Fe, NM, Nov. 6-9, 2002.

\bibitem{Hnat}
B. Hnat, S.C. Chapman, and G. Rowlands, {\it Intermittency,
scaling and the Fokker-Planck approach to fluctuations of the
solar wind bulk plasma parameters as seen by WIND},
physics/0211080 (2002).

\bibitem{Laval}
J.-P. Laval, B. Dubrulle, and S. Nazarenko, Phys. Fluids {\bf 13},
1995 (2001); {\it Non-locality and intermittency in 3D
turbulence}, physics/0101036 (2001).

\bibitem{Laval2}
J.-P. Laval, B. Dubrulle, and J.C. McWilliams,
Phys. Fluids {\bf 15}, 1327 (2003).

\bibitem{Muzy}
J.F. Muzy and E. Bacry, {\it Multifractal stationary random
measures and multifractal random walks with log-infinitely
divisible scaling laws}, cond-mat/0206202.

\bibitem{Pinton}
N. Mordant, P. Metz, O. Michel, and J.-F. Pinton, {\it Measurement
of Lagrangian velocity in fully developed turbulence},
physics/0103084; Phys. Rev. Lett. {\bf 87}, 214501 (2001).

\bibitem{Friedrich}
Ch. Renner, J. Peinke, and R. Friedrich, {\it On the interaction
between velocity increment and energy dissipation in the turbulent
cascade}, physics/0211121 (2002).
 Ch. Renner, J. Peinke, R. Friedrich, O. Chanal, and B. Chabaud,
 Phys. Rev. Lett. {\bf 89}, 124502 (2002).

\bibitem{Chapman}
S.C. Chapman, G. Rowlands, and N.W. Watkins, {\it The origin of
universal fluctuations in correlated systems: explicit calculation
for an intermittent turbulent cascade}, cond-mat/0302624 (2003).


\bibitem{Portelli}
B. Portelli, P.C.W. Holdsworth, and J.-F. Pinton, {\it
Intermittency and non-Gaussian fluctuations of the global energy
transfer in fully developed turbulence}, cond-mat/0112503 (2001).

\bibitem{Nazarenko}
S. Nazarenko, N. K.-R. Kevlahan, and B. Dubrulle,
Physica D {\bf 139}, 158 (2000).

\bibitem{Nakao}
H. Nakao, {\it Asymptotic power law of moments in a random
multiplicative process with weak additive noise}, cond-mat/9802030
(1998).

\bibitem{Kuramoto}
Y. Kuramoto and H. Nakao, Phys. Rev. Lett. {\bf 76}, 4352 (1996);
{\bf 78}, 4039 (1997).





\end{thebibliography}
\end{document}